\documentclass[a4paper, 11pt]{article}
\usepackage{jheppub}

\usepackage[
    colorlinks=true, linktocpage=true,
    linkcolor=blue, citecolor=blue, 
    urlcolor=blue]{hyperref}

\usepackage[T1]{fontenc}
\usepackage{amsmath}
\usepackage{graphicx}
\usepackage{epstopdf}
\usepackage{cancel}
\usepackage{slashed}
\usepackage{booktabs}
\usepackage{placeins}

\allowdisplaybreaks[1]

\title{One-loop QCD amplitudes for vector-boson-fusion Higgs-boson production to 
higher orders in \texorpdfstring{\boldmath $\epsilon$}{ep}}

\author[a]{Konstantin Asteriadis,}
\author[a]{Linus G\"otzfried,}
\author[a]{Andreas von Manteuffel}

\affiliation[a]{Institute for Theoretical Physics, University of Regensburg, 
93053 Regensburg, Germany}

\emailAdd{konstantin.asteriadis@ur.de}
\emailAdd{linus.goetzfried@ur.de}
\emailAdd{manteuffel@ur.de}

\abstract{We consider one-loop QCD corrections to Higgs-boson production via 
vector-boson fusion at hadron colliders. As an input for future exact NNLO 
computations, we derive the one-loop helicity amplitudes to higher orders in the 
dimensional regulator. We present a direct calculation of the amplitudes with 
spinor-helicity techniques and different $\gamma_5$ schemes. In addition, we 
compare this approach to a projector-based calculation. In order to derive 
analytical results to higher orders in $\epsilon$, we calculate the pentagon 
master integrals by the method of $\epsilon$-factorised differential equations 
and express them in terms of an algebraically independent set of basis 
functions.}

\begin{document}

\maketitle
\flushbottom


\section{Introduction}
\label{sec:introduction}

Higgs-boson production in vector-boson fusion (VBF) has a distinctive signature, 
most importantly the presence of two hard jets, one along each beam axis. 
Application of strict fiducial cuts isolates VBF from identical final-state 
processes such as Higgs-Strahlung Higgs-boson production (VH) with hadronic 
decay. Within this fiducial region, VBF has been extensively studied.

The fiducial cross section is dominated by so-called factorisable contributions, 
which involve no colour exchange between colliding partons. For these, 
differential predictions include up to next-to-leading order (NLO) 
electroweak~\cite{Ciccolini:2007ec, Figy:2010ct}, and up to 
next-to-next-to-leading order (NNLO) quantum chromodynamics 
(QCD)~\cite{Figy:2003nv, Figy:2004pt, Berger:2004pca, Bolzoni:2010xr, 
Bolzoni:2011cu, Cacciari:2015jma, Cruz-Martinez:2018rod} corrections. Inclusive 
cross sections have been computed to next-to-next-to-next-to-leading order 
(N$^3$LO) QCD~\cite{Dreyer:2016oyx}, and current efforts push for differential 
predictions to reach the same level of accuracy.

In contrast, non-factorisable corrections are in general small, suppressed by 
colour and, in some cases, additionally by kinematics. Non-factorisable 
double-virtual QCD amplitudes, required for NNLO QCD accuracy, have been 
computed in the eikonal approximation, including next-to-leading 
contributions~\cite{Liu:2019tuy, Long:2023mvc}. This approximation is valid 
inside the fiducial VBF region, where it provides sufficient accuracy for 
phenomenology even at N$^3$LO QCD precision~\cite{Asteriadis:2023nyl, 
Long:2023mvc}.

Despite these impressive results, our understanding might not be sufficient. The 
expected experimental precision of the high-luminosity phase of the Large Hadron 
Collider and improved analysis techniques allow for relaxing fiducial cuts and 
exploring broader kinematic ranges. Standard template cross sections for the 
HL-LHC combine VBF and VH Higgs-boson production channels and partially drop 
strict VBF cuts~\cite{Berger:2019wnu}. Outside the traditional fiducial volume, 
the validity of current NNLO QCD predictions and separation from VH is 
uncertain~\cite{Dreyer:2020urf}, and the currently unknown exact 
non-factorisable virtual contributions prevent a definitive study.

In this paper, we provide a first step towards exact non-factorisable virtual 
corrections at NNLO by deriving the one-loop helicity amplitudes to higher 
orders in the dimensional regulator $\epsilon$ (see \cite{Goetzfried:2026pbhb} for first two-loop integral results). At $\mathcal{O}(\alpha_s^2)$, 
due to the presence of infrared poles, the original one-loop computations to 
$\mathcal{O}(\epsilon^0)$ \cite{Figy:2003nv, Figy:2004pt, Berger:2004pca} need 
to be extended. Moreover, since higher orders in $\epsilon$ require a new set of 
genuine five-point one-loop Feynman integrals, the amplitudes can also not be 
computed using automatic one-loop providers.

A standard way to compute such helicity amplitudes in dimensional regularisation 
uses projections of the amplitude on physical Lorentz structures. It is not 
straightforward to use more direct methods based on the spinor-helicity 
formalism, since the latter is inherently four dimensional. In this paper, we 
apply the spinor-helicity formalism to dimensionally regulated amplitudes and 
discuss different variants of this scheme. In particular, we compute the 
amplitudes multiple times using projectors or spinor-helicity techniques with 
different prescriptions for $\gamma_5$, and compare the different methods.

We establish our notation in section~\ref{sec:notation}, and provide a reference 
calculation using the established projector method in 
section~\ref{sec:projectors}. In section~\ref{sec:helamp:anticom} we investigate 
spinor-helicity methods in dimensional regularisation with anticommuting 
$\gamma_5$, and in section~\ref{sec:helamp:thooft-veltman} with a 
Breitenlohner-Maison~/~'t~Hooft-Veltman (BMHV) $\gamma_5$ scheme. The 
computation of the required master integrals is discussed in 
section~\ref{sec:master-integrals}, followed by ultraviolet renormalisation and 
infrared subtraction in section~\ref{sec:uv-ir}. In section~\ref{sec:numeric} we 
present numerical results and validate against existing work. We conclude in 
section~\ref{sec:conclusion}.


\section{Notation}
\label{sec:notation}

\begin{figure}[t]
    \centering
    \includegraphics[height=60pt]{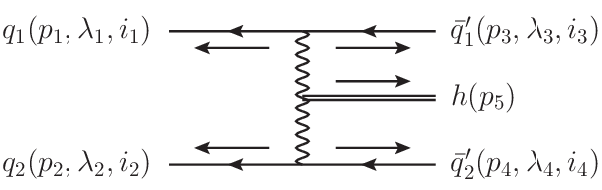}
    \caption{Conventions for (anti-)quark flavours $q_{j = 1,2}^{(\prime)}$, 
    momenta $p_{j=1 \dots 5}$, helicities $\lambda_{j = 1 \dots 4}$, and colour 
    indices $i_{j = 1 \dots 4}$ used throughout the discussion.
    All particles and momenta are chosen outgoing.}
    \label{fig:born}
\end{figure}

We consider Higgs-boson production through vector-boson fusion of 
distinguishable quarks. Taking all particles and momenta outgoing, the process 
of interest is
\begin{align} 
    \label{eq:notation:proc}
    0 \rightarrow q_1(p_1) + q_2(p_2) + \bar{q}_1^\prime(p_3) 
        + \bar{q}_2^\prime(p_4) + h(p_5) \,.
\end{align}
In figure~\ref{fig:born}, we illustrate these conventions for the Born-level 
topology and include also helicity and colour indices. We compute amplitudes as 
a perturbative series in the bare strong coupling constant 
$\alpha_{s,\textrm{b}} = g_{s,\textrm{b}}^2/(4\pi)$,
\begin{align}
\label{eq:ampl:perturbationseries}
    \mathcal{A} = \mathcal{A}^{(0)} 
        + \frac{\alpha_{s,\textrm{b}}}{2\pi} \mathcal{A}^{(1)} 
        + \mathcal{O}(\alpha_{s,\textrm{b}}^2) \, .
\end{align}
Ultraviolet (UV) and infrared (IR) divergences are regulated using non-integer 
space-time dimension $d = 4 - 2 \epsilon$.

To keep results generic and valid for all quark flavours and vector-boson types, 
we use the quark-weak-vector-boson vertex
\begin{align}
    \label{eq:feynman:qqv}
    \hspace{-10pt}
    \vcenter{\hbox{\includegraphics[height=50pt]{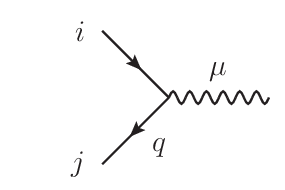}}} \quad 
        = \quad \mathrm{i}e \, \delta_{ij} \, \gamma^\mu \, 
        (g_{q}^+\,\textrm{P}_++g_{q}^-\,\textrm{P}_-)\, ,
\end{align}
where $e = \sqrt{4\pi\alpha}$ is the positron charge, $g_{q}^\pm$ are right- and 
left-handed couplings to a quark of type $q$ and the projectors are defined as
\begin{align}
    \textrm{P}_{\pm} \equiv \frac{1 \pm \gamma_5}{2} \, .
\end{align}
For the Higgs-vector-boson coupling, we use
\begin{align}
    \label{eq:feynman:vvh}
    \vcenter{\hbox{\includegraphics[height=50pt]{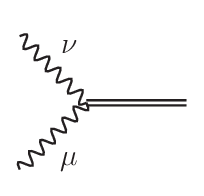}}} \quad 
        = \quad \mathrm{i} \, g_\textrm{vvh}\,g^{\mu\nu} \, .
\end{align}
The explicit forms of the coupling constants $g_\textrm{vvh}$ and $g_{q}^\pm$ 
are of no relevance for the current discussion.

Helicity amplitudes are written as 
$\mathcal{A}_{\lambda_1\lambda_2\lambda_3\lambda_4}$, where the parton with 
momentum $p_i$ carries helicity $\lambda_i\in\{+,-\}$, cf.\ 
figure~\ref{fig:born}. Because we treat the quarks as massless, helicity is 
conserved, leaving only four non-vanishing configurations, $\mathcal{A}_{--++}$, 
$\mathcal{A}_{-++-}$, $\mathcal{A}_{+--+}$ and $\mathcal{A}_{++--}$. In what 
follows, we omit the final two helicity labels for brevity, 
$\mathcal{A}_{\lambda_1 \lambda_2} \equiv \mathcal{A}_{\lambda_1 \lambda_2 
\lambda_3 \lambda_4}$.

\begin{figure}[t]
    \centering
    \includegraphics[height=80pt]{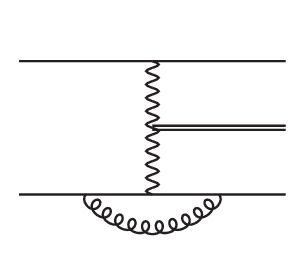} \hspace{30pt}
    \includegraphics[height=80pt]{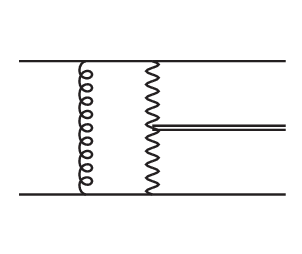}
    \includegraphics[height=80pt]{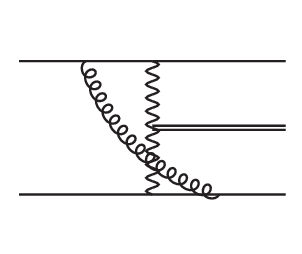} \\
    \includegraphics[height=80pt]{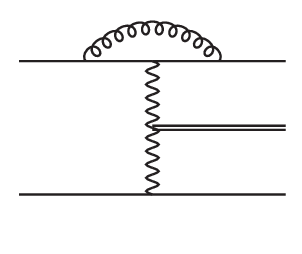} \hspace{30pt}
    \includegraphics[height=80pt]{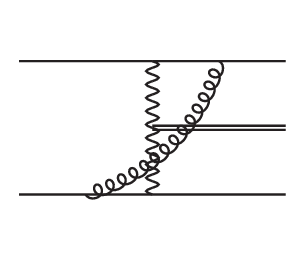}
    \includegraphics[height=80pt]{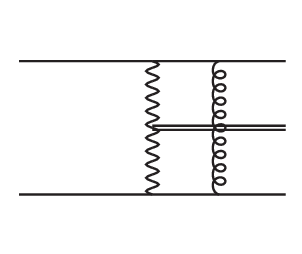} \\[-15pt]
    (a) \hspace{161pt} (b) \hspace{44pt}
    \caption{Factorisable (a) and non-factorisable (b) diagrams contributing 
    to the one-loop amplitude.
    }
    \label{fig:1loop}
\end{figure}

We work with colour-stripped amplitudes and factor out common coupling 
constants. At Born level, we write
\begin{align}
\label{eq:ampl:born:A}
    \mathcal{A}^{(0)}_{\lambda_1\lambda_2} = e^2 \, g_\textrm{vvh} \, 
        g_{q_1}^{\lambda_1} \, g_{q_2}^{\lambda_2} \ \delta_{i_1 i_3} 
        \delta_{i_2 i_4} \ A^{(0)}_{\lambda_1\lambda_2} \, ,
\end{align}
where $i_k$ is the colour index of parton $k$, cf. figure~\ref{fig:born}. 
One-loop corrections are shown in figure~\ref{fig:1loop}. Depending on whether 
or not there is colour exchange between the two incoming quarks, they have 
different colour structures. Based on these structures, they are sorted into 
individually gauge-invariant factorisable (f) and non-factorisable (nf) 
contributions,
\begin{align}
    \label{eq:ampl}
    \mathcal{A}^{(1)}_{\lambda_1\lambda_2} = e^2 \,
    g_\textrm{vvh} \, g_{q_1}^{\lambda_1} \, g_{q_2}^{\lambda_2} 
    \Big(C_f \, \delta_{i_1 i_3}\delta_{i_2 i_4} \, 
    A^{(1),\textrm{f}}_{\lambda_1\lambda_2} + T^a_{i_1 i_3} \, 
    T^a_{i_2 i_4} \, A^{(1),\textrm{nf}}_{\lambda_1\lambda_2} \Big) \, ,
\end{align}
where 
\begin{align}
    C_f = \frac{N_c^2 - 1}{2\,N_c} \, , \quad T^a_{i_1 i_3} \, T^a_{i_2 i_4} 
        = \frac{1}{2} \bigg(\delta_{i_1 i_4} \delta_{i_2 i_3} - \frac{1}{N_c} 
        \delta_{i_1 i_3} \delta_{i_2 i_4} \bigg) \, ,
\end{align}
and $N_c = 3$ is the number of quark colour charges.

The helicity amplitudes depend on spinor structures carrying the little-group 
weight. The Lorentz invariant part can be parametrised by a choice of six 
independent Mandelstam invariants, $s_{ij} \equiv (p_i + p_j)^2$, and the mass 
square $m_\textrm{v}^2$ of the vector boson. An important quantity for 
five-point kinematics is the pseudoscalar
\begin{align}
    \label{eq:pseudoscalar}
    \mathrm{tr}_5 \equiv \mathrm{tr}(\slashed{p}_1 \slashed{p}_2
    \slashed{p}_3 \slashed{p}_4 \gamma_5) = [12] \langle 23\rangle [34] 
    \langle 41\rangle - \langle 12\rangle [23] \langle 34\rangle [41] \, ,
\end{align}
where $\langle i j \rangle$ and $[ij]$ are the usual spinor products.
Note that $\mathrm{tr}_5^2=16\,\Delta_5$, where
\begin{align}
    \Delta_5 &\equiv 
        \begin{vmatrix}
            p_1 \cdot p_1 & \dots & p_1 \cdot p_4 \\
            \vdots & \ddots & \vdots \\
            p_4 \cdot p_1 & \dots & p_4 \cdot p_4
        \end{vmatrix} 
        = \frac{1}{16} \big(s_{14}^2 s_{23}^2 + (s_{13} s_{24} - s_{12} 
        s_{34})^2 - 2 s_{14}s_{23} (s_{13} s_{24} + s_{12} s_{34}) \big) 
        \, , \nonumber\\[-10pt]
    \label{eq:gramdet}
\end{align}
is the five-point Gram determinant.


\section{Projector method}
\label{sec:projectors}

We compute the VBF amplitude with different methods and we start with the 
established projector method. Amplitudes are decomposed into a linear 
combination of basis tensors multiplying Lorentz-scalar form factors. The latter 
are obtained by projecting the amplitude on the tensor basis.
We employ polarisation sums in $d$ dimensions,
\begin{align}
    \sum_\lambda u_\lambda(p)\bar{u}_\lambda(p) 
        = \sum_\lambda v_\lambda(p)\bar{v}_\lambda(p)
        \equiv \slashed{p} = p_\mu \gamma^\mu,
\end{align}
with $\gamma^\mu$ a $d$-dimensional $\gamma$-matrix. For ease of computability, 
we choose an anticommuting $\gamma_5$-scheme~\cite{Chanowitz:1979zu}, which is 
expected to be an effectively valid choice for the process and loop order under 
consideration. In particular, there are no anomalous contributions involving 
closed fermion loops and $\gamma_5$ enters only on the external quark lines. 
Within such a scheme, we may even eliminate all explicit occurrences of 
$\gamma_5$ from the computation (compare e.g. ref.~\cite{Gehrmann:2015ora}). 
Indeed, since quarks are taken to be massless and both QCD and electroweak 
fermion couplings include a single gamma matrix, all spinor chains occurring in 
the amplitude are of the form
\begin{align}
    \label{eq:proj:spinorchain}
    g_{q}^\pm \, \bar{u} \gamma^{\mu_1} \dots \gamma^{\mu_{2m+1}} 
        \textrm{P}_{\pm} \gamma^{\nu_1} \dots \gamma^{\nu_{2n}} v 
        = g_{q}^\pm \, \bar{u} \gamma^{\mu_1} \dots \gamma^{\mu_{2m+1}} 
        \gamma^{\nu_1} \dots \gamma^{\nu_{2n}} \textrm{P}_{\pm} v \, ,
\end{align}
where we used the anticommuting property of $\gamma_5$ to move the 
$d$-dimensional projector $\textrm{P}_\pm$ to the very right. The projectors are 
defined to have the same action on spinors as their four-dimensional 
equivalents, and hence eq.~\eqref{eq:proj:spinorchain} is zero for one helicity 
choice of $v$ and equal to $g_{q}^\pm \, \bar{u} \gamma^{\mu_1} \dots 
\gamma^{\mu_{2m+1}} \gamma^{\nu_1} \dots \gamma^{\nu_{2n}} v$ otherwise. Since 
apart from the coupling constant, the non-vanishing contribution takes on the 
same form for all different helicity combinations, we can define a ``proxy'' 
amplitude $\mathcal{A}^\prime$ analogously to $\mathcal{A}$, but with the 
Feynman rule for the quark-weak-vector-boson vertex replaced by 
\begin{align}
    \hspace{-10pt}
    \vcenter{\hbox{\includegraphics[height=50pt]{figs/qqv.eps}}} \quad = \quad
    \delta_{ij} \, \gamma^\mu \, .
\end{align}
For the helicity amplitudes $\mathcal{A}_{\lambda_1\lambda_2}^\textrm{pr}$ 
computed in the projector approach, we then have
\begin{align}
    \label{eq:proj:proxy}
    \mathcal{A}_{\lambda_1\lambda_2}^\textrm{pr} = g_{q_1}^{\lambda_1} 
    g_{q_2}^{\lambda_2} \, \mathcal{A}_{\lambda_1\lambda_2}^\prime 
    \equiv g_{q_1}^{\lambda_1} g_{q_2}^{\lambda_2} g_{\textrm{vvh}} \, 
    A_{\lambda_1\lambda_2}^\prime \, , 
\end{align}
for all helicity combinations $\lambda\in\{--,-+,+-,++\}$, and it is sufficient 
to compute helicity amplitudes for $A^\prime$. 

One can express the amplitude $A^\prime$ as a linear combination of basis 
tensors $T_i$ with Lorentz-scalar form factors $F_i$ as coefficients,
\begin{align}
\label{eq:proj:FFdecomp}
    A^\prime = \sum_i F_i\, T_i \, .
\end{align}
In $d$ dimensions, the number of possible linearly independent tensors involving 
two external fermion lines increases with the loop order. Ultimately, we are 
interested in the limit $d \to 4$, and in four dimensions, only a small finite 
set of tensors are linearly independent. One can systematically decouple the 
additional evanescent tensors in $d$ space-time dimensions, such that after UV 
and IR subtraction the $\mathcal{O}(\epsilon^0)$ part of the finite remainder 
receives no contribution from the evanescent structures~\cite{Peraro:2020sfm}. 
In practice, one can obtain the $\mathcal{O}(\epsilon^0)$ part of the finite 
remainder by just ignoring the evanescent tensors right from the start and 
restricting the sum in \eqref{eq:proj:FFdecomp} to a set of tensors $T_i$ which 
form a basis in four dimensions. Here, we choose
\begin{align}
\label{eq:proj:tensors}
\begin{aligned}[b]
    T_1 &= \bar{u}(p_1) \slashed{p}_2 v(p_3) \, 
        \bar{u}(p_2) \slashed{p}_1 v(p_4) \, , & \quad 
    T_3 &= \bar{u}(p_1) \slashed{p}_4 v(p_3) \, 
        \bar{u}(p_2) \slashed{p}_1 v(p_4) \, , & \\
    T_2 &= \bar{u}(p_1) \slashed{p}_2 v(p_3) \, 
        \bar{u}(p_2) \slashed{p}_3 v(p_4) \, , & \quad
    T_4 &= \bar{u}(p_1) \slashed{p}_4 v(p_3) \, 
        \bar{u}(p_2) \slashed{p}_3 v(p_4) \, . & 
\end{aligned}
\end{align}
To obtain the form factors $F_i$, we construct projectors $P_i$
such that 
\begin{align}
    F_i=\sum_{\lambda} P_i\,A^\prime\,.
\end{align}
This implies
\begin{align}
    \sum_{\lambda}P_i\,T_j \overset{!}{=} \delta_{ij} \quad \Leftrightarrow \quad 
        P_i = \sum_j (M^{-1})_{ij} \, T_j^\dagger \quad \textnormal{with} \quad 
        M_{ij} \equiv \sum_{\lambda} T_i^\dagger T_j \, ,
\end{align}
and the projectors are determined by inverting a $4\times 4$-matrix. The 
projectors are regular in the limit $\epsilon \to 0$ and the details of the 
higher order terms in $\epsilon$ will ultimately not matter at the level of the 
$\epsilon^0$ part of the finite remainder. It is important, however, to stick to 
one specific choice of the projectors at all loop orders. In particular, this 
will ensure the construction of consistent IR subtraction terms in that scheme.

We obtain helicity amplitudes from eq.~\eqref{eq:proj:FFdecomp} by fixing the 
helicities in $T_{1\dots 4}$ in all possible ways, i.e.\ choosing specific 
polarisations of the $\bar{u}$- and $v$-spinors appearing in 
eq.~\eqref{eq:proj:tensors}, and evaluating the tensors in four space-time 
dimensions. We introduce spinor factors
\begin{align}
\label{eq:proj:spinorfactors}
\begin{aligned}[b]
    \Phi_{--} &\equiv 2\langle 12 \rangle[34]\,, \quad & \Phi_{-+} &\equiv 
        -2\langle 14\rangle[23]\, , \\
    \Phi_{++} &\equiv 2[12]\langle34\rangle = \Phi_{--}^\star\,,\quad & 
        \Phi_{+-} &\equiv -2[14]\langle23\rangle = \Phi_{-+}^\star \, ,
\end{aligned}
\end{align}
that correspond to the physical helicity structures at Born level and that carry 
the little-group weight of each helicity amplitude. We choose the normalisation 
in eq.~\eqref{eq:proj:spinorfactors} such that 
\begin{align}
    A_\lambda^{\prime(0)} =  \frac{\Phi_\lambda}{(s_{13} - m_\mathrm{v}^2)(s_{24} 
        - m_\mathrm{v}^2)} + \mathcal{O}(\epsilon)\, .
\end{align}
After dividing those out from the right-hand side of 
eq.~\eqref{eq:proj:FFdecomp}, the remaining part can be expressed in terms of 
Lorentz invariants and $\textrm{tr}_5$. We obtain
\begin{align}
        A^\prime_{--} &= -\frac{\Phi_{--}}{2} \bigg(s_{12} F_1+s_{23}F_2
            +s_{14}F_3+s_{34}F_4 \nonumber\\
        &\hspace{40pt}-\frac{1}{2} \, (s_{12} s_{34} - s_{14} s_{23}
            + s_{13}s_{24})\,\bigg(\frac{F_1}{s_{34}}+\frac{F_4}{s_{12}}\bigg)
            - \frac{1}{2}\,\mathrm{tr}_5\,\bigg(\frac{F_1}{s_{34}} 
            - \frac{F_4}{s_{12}}\bigg)\!\bigg)\, ,
    \label{eq:ampl:ff} \\
        A^\prime_{-+} &= -\frac{\Phi_{-+}}{2} \bigg(s_{12} F_1+s_{23}F_2
            +s_{14}F_3+s_{34}F_4 \nonumber\\
        &\hspace{40pt}-\frac{1}{2}\,(s_{24}s_{13}-s_{12}s_{34}
            +s_{23}s_{14})\,\bigg(\frac{F_2}{s_{14}}+\frac{F_3}{s_{23}}\bigg)
            -\frac{1}{2}\,\mathrm{tr}_5\,\bigg(\frac{F_2}{s_{14}}
            -\frac{F_3}{s_{23}}\bigg)\!\bigg)\, , \\
    A^\prime_{+-} &= A^\prime_{-+}\Big|_{\langle\,\,\rangle
        \leftrightarrow[\,\,]}\, ,\\
    \label{eq:ampl:ff4}
    A^\prime_{++} &= A^\prime_{--}\Big|_{\langle\,\,\rangle
        \leftrightarrow[\,\,]}\, .
\end{align}
The operation $\langle\,\,\rangle\leftrightarrow[\,\,]$ acts on the amplitudes 
via replacements 
\begin{align}
\label{eq:proj:conjugate}
    \Phi_{--}\leftrightarrow\Phi_{++}\,,\quad \Phi_{-+}\leftrightarrow\Phi_{+-} 
    \, , \quad\textnormal{and}\quad \mathrm{tr}_5\leftrightarrow-\mathrm{tr}_5\,,
\end{align}
where by construction of the amplitude $\mathcal{A}^\prime$, the form factors 
depend only on Lorentz invariants and not on $\mathrm{tr}_5$.

For the computation of the form factors at tree and one-loop level, we followed an 
established toolchain. The relevant Feynman diagrams were generated using 
\texttt{Qgraf}~\cite{Nogueira:1991ex}, see figure~\ref{fig:1loop}.
We applied Lorentz projectors $P_i$, performed colour and Lorentz 
algebra with an in-house collection of \texttt{FORM}~\cite{Kuipers:2012rf, 
Davies:2026cci} and \texttt{Mathematica} scripts. Remaining scalar integrals are 
reduced to a set of master integrals using integration-by-parts (IBP) 
techniques. For 
details regarding the calculation of master integrals, see 
section~\ref{sec:master-integrals}.


\section{Spinor-helicity method in dimensional regularisation}
\label{sec:helamp}

In this section, we describe how helicity amplitudes can be derived in an 
alternative way to the projector method described above, using spinor-helicity 
techniques.
We start with a discussion of the tree-level amplitude, which already 
illustrates the general approach. The relevant part of the Born amplitude is 
given by
\begin{align}
    \label{eq:ampl:born}
    \mathcal{A}^{(0)} &\sim \bar{u}(p_1) \gamma^\mu (g_{q_1}^- \textrm{P}_- 
        + g_{q_1}^+ \textrm{P}_+) v(p_3) \times \bar{u}(p_2) \gamma_\mu (g_{q_2}^- \textrm{P}_- 
        + g_{q_2}^+ \textrm{P}_+) v(p_4) \, ,
\end{align}
where we omitted Higgs-weak-boson couplings and propagators.
In physical four dimensional space-time, computing amplitudes for defined external 
polarisations is routinely done using spinor-helicity methods. As an example, we 
consider left-handed $q_1$ and $q_2$, use $\bar{u}_-(p_i) \equiv \langle i |$ 
and $v_+(p_i) \equiv | i ]$, and find
\begin{align}
    \label{eq:ampl:born:ll:4d}
    A^{(0),\textrm{4D}}_{--}  \sim  \langle 1 | \gamma^\mu |3]  \langle 2 
    | \gamma_\mu |4] = -2 \langle 12 \rangle [34] \, ,
\end{align}
where we used a Fierz identity,
\begin{align}
    \label{eq:fierz:4d}
    \langle p | \gamma^\mu |q]  \langle k | \gamma_\mu |l] 
        = 2 \langle pk \rangle [lq] \, , 
\end{align}
to sum over the contracted Lorentz index $\mu$. 

We would like to do something similar in dimensional regularisation. In 
particular, we aim at schemes, where external non-singular particles are kept 
four-dimensional. It is customary and sufficient to keep Dirac space quasi-four 
dimensional and use an identity matrix $I$ with $\mathrm{tr}(I) = 
4$~\cite{tHooft:1972tcz}. However, in general it is not clear how 
$d$-dimensional Dirac matrices $\gamma^\mu$ and projectors $\textrm{P}_{\pm}$ 
should act on four-dimensional helicity spinors. By introducing additional 
projectors,\footnote{We note that, in general we cannot replace an expression as 
e.g.\ $\textrm{P}_- | i ]$ simply by $| i ]$, as it would be the case in four 
dimensions.}
\begin{align}
    u_-(p_i) = v_+(p_i) = \textrm{P}_- | i ] \, , \quad 
    u_+(p_i) = v_-(p_i) = \textrm{P}_+ | i \rangle \, , \nonumber\\
    \bar{u}_-(p_i) = \bar{v}_+(p_i) = \langle i| \textrm{P}_+ \, , \quad 
    \bar{u}_+(p_i) = \bar{v}_-(p_i) = [i| \textrm{P}_- \, , 
    \label{eq:spinors:ddim}
\end{align}
information on the helicity is now explicit and we do not need to act with 
$\gamma^\mu$ directly on helicity spinors to include these effects correctly. It 
is sufficient to define commutators and anticommutators for $d$-dimensional 
$\gamma_5$ and $\gamma_\mu$, i.e.\ choose a $\gamma_5$-scheme.

In the following, we will consider two schemes for $\gamma_5$, anticommuting and 
Breitenlohner-Maison~/~'t Hooft-Veltman. Using left-handed spinors from eq.~\eqref{eq:spinors:ddim} in 
the amplitude~\eqref{eq:ampl:born}, we find
\begin{align}
    A^{(0)}_{--} &\sim \langle 1 | \textrm{P}_+ \gamma^\mu (g_{q_1}^- \textrm{P}_- + g_{q_1}^+ \textrm{P}_+) 
        \textrm{P}_- | 3] \times\langle 2 | \textrm{P}_+\gamma_\mu (g_{q_2}^- \textrm{P}_- + g_{q_2}^+ \textrm{P}_+) 
        \textrm{P}_- | 4] \nonumber\\
    &= g_{q_1}^- g_{q_2}^-  \langle 1 | \textrm{P}_+ \gamma^\mu  \textrm{P}_- | 3]  \langle 2 | 
        \textrm{P}_+\gamma_\mu  \textrm{P}_- | 4] \, ,
    \label{eq:ampl:born:ddim}
\end{align}
where we made use of $\textrm{P}_i \textrm{P}_j = \delta_{ij} \textrm{P}_i$, 
which is valid in both considered $\gamma_5$-schemes. Similarly to the 
four-dimensional case, we would like to remove the contracted Lorentz index 
$\mu$, but in general, the Fierz identity eq.~\eqref{eq:fierz:4d} is not valid 
in $d$-dimensions. 

To understand how this identity can be extended to $d$-dimensions, we note that 
in four dimensions it can be computed by inserting unity expressed as
\begin{align}
    \label{eq:unity}
    1 = \frac{\langle 3 | \slashed{k} | 2]}{\langle 3 | \slashed{k} | 2]}
\end{align}
into the left-hand side of 
eq.~\eqref{eq:fierz:4d},
\begin{align}
    \label{eq:ampl:unity_inserted}
    \langle 1 | \gamma^\mu | 3 ]  \langle 2 | \gamma_\mu | 4 ] 
        =  \frac{\langle 1 | \gamma^\mu | 3 ] \langle 3 | \slashed{k} | 2] 
            \langle 2 | \gamma_\mu | 4 ]}{\langle 3 | \slashed{k} | 2]} 
        = \frac{\langle 1 | \gamma^\mu \slashed{3}  \slashed{k} \slashed{2} 
            \gamma_\mu | 4 ]}{\langle 3 | \slashed{k} | 2]} \, .
\end{align}
Here, $\slashed{k} = k^\mu \gamma_\mu$ is an arbitrary four-dimensional momentum 
(that is linearly independent from $p_2$ and $p_3$) contracted with a Dirac 
gamma matrix. In the last step we used the completeness relation in four 
dimensions,
\begin{align}
    \label{eq:completeness:4d}
    \slashed{p}_i = |i\rangle[i| + |i]\langle i| \, ,
\end{align}
to join spinor chains in the numerator. At this point, by repeated application 
of the anticommutator relation $\{\gamma^\mu,\gamma^\nu\}=2\,g^{\mu\nu}$, we may 
straightforwardly remove the contracted Lorentz index $\mu$ and arrive after 
simplifications at the right-hand side of eq.~\eqref{eq:fierz:4d}. In the 
following  sections we discuss this procedure in dimensional regularisation and 
begin with an anticommuting $\gamma_5$-scheme.


\subsection{Anticommuting \texorpdfstring{\boldmath $\gamma_5$}{gamma 5}}
\label{sec:helamp:anticom}

In an anticommuting $\gamma_5$-scheme, all Dirac matrices $\gamma^\mu$ are 
$d$-dimensional. We use
\begin{align}
    \label{eq:gamma_norm}
    \gamma^\mu \gamma_\mu = d \, I\, , \quad \gamma_5^2 = I \, , 
\end{align}
where we recall that $I$ is the identity matrix in Dirac spinor space.
We preserve Clifford algebra and anticommutation with $\gamma_5$ 
\begin{align}
    \label{eq:anticom}
    \{\gamma^\mu, \gamma^\nu\} &= 2g^{\mu\nu} \, I \, , \quad 
        \{\gamma^\mu, \gamma_5\} = 0 \, .
\end{align}
We can still introduce unity~\eqref{eq:unity} into the amplitude~\eqref{eq:ampl:born:ddim}, 
but, following eq.~\eqref{eq:spinors:ddim}, we 
write it using explicit projectors between external physical spinors and 
$d$-dimensional objects in Dirac space,
\begin{align}
    \label{eq:ampl:anti}
    A^{(0),\textrm{ac}}_{--} &\sim \frac{  \langle 1 | \textrm{P}_+ \gamma^\mu  
        \textrm{P}_- | 3] \langle 3 | \textrm{P}_+ \slashed{k} \textrm{P}_- | 2] \langle 2 | 
        \textrm{P}_+\gamma_\mu  \textrm{P}_- | 4]}{\langle 3 | \textrm{P}_+ \slashed{k} \textrm{P}_- | 2]} \, ,
\end{align}
where the superscript ``ac'' indicates a $d$-dimensional amplitude computed in 
our anticommuting $\gamma_5$ scheme. To continue, we could use the 
four-dimensional completeness relation eq.~\eqref{eq:completeness:4d} to combine 
spinor chains, but then we would need to prescribe an algebra including both 
four- and $d$-dimensional $\gamma$-matrices. One such algebra is the 
BMHV $\gamma_5$-scheme and we discuss it in the next section. In 
the current case, with anticommuting $\gamma_5$, we simply \emph{prescribe}
\begin{align}
    \label{eq:completeness:anti}
    \slashed{p}_i \equiv \textrm{P}_+|i\rangle[i|\textrm{P}_- + \textrm{P}_-|i]\langle i|\textrm{P}_+ \, .
\end{align}
Note that the left-hand side of eq.~\eqref{eq:completeness:anti} includes 
unphysical helicity configurations even for $4$-dimensional $p_i$, since 
$\slashed{p}_i=p_i^\mu\gamma_\mu$ contains genuinely $d$-dimensional 
$\gamma$-matrices. On the right-hand side, we have dropped them by introducing 
projectors. This introduces evanescent $\mathcal{O}(\epsilon)$ contributions 
throughout the computation, which should cancel in finite physical results, 
provided the prescription eq.~\eqref{eq:completeness:anti} is applied 
consistently at all perturbative orders. We confirm that this cancellation 
indeed occurs for this computation, and we elaborate on the exact requirements 
for this consistency below.
 
Using the prescription~\eqref{eq:completeness:anti} in 
eq.~\eqref{eq:ampl:anti} and assuming $k^2 = 0$, we find
\begin{align} 
    A^{(0),\textrm{ac}}_{--} &\sim \frac{\langle 1| \textrm{P}_+ \gamma^\mu \textrm{P}_-| 3] 
        \langle 3 | \textrm{P}_+ \slashed{k} \textrm{P}_- | 2] \langle 2 | \textrm{P}_+ \gamma_\mu \textrm{P}_- |4]}
        {\langle 3 | \textrm{P}_+ \slashed{k} \textrm{P}_- | 2]} 
    = \frac{\langle 1| \textrm{P}_+  \gamma^\mu \slashed{3} \slashed{k} \slashed{2} 
        \gamma_\mu \textrm{P}_- |4]}{\langle 3 | \textrm{P}_+ \slashed{k} \textrm{P}_- | 2]}  \nonumber\\
    &= -2 \langle 1 2 \rangle [34] + 2 \epsilon \frac{\langle k 2 \rangle 
        \langle 1 3 \rangle [k3][24]}{\langle k 3 \rangle [k2]} \, .
    \label{eq:ampl:anti:ll}
\end{align}
In the last step we used eqs.~(\ref{eq:gamma_norm}~-~\ref{eq:anticom}) to remove 
the contracted Lorentz index $\mu$. At this point, spinor chains only contain 
physical momenta and we finished the computation using four-dimensional 
identities. For instance, we used Schouten identities
\begin{align}
        [ij][kl] + [ik][lj] + [il][jk] &= 0 \, , \nonumber\\
        \langle ij\rangle \langle kl\rangle + \langle ik\rangle \langle lj
            \rangle + \langle il\rangle \langle jk\rangle &= 0 \, ,
    \label{eq:schouten}
\end{align}
to move $\mathcal{O}(\epsilon)$ contributions from the first to the second term.

The first term on the right-hand side of eq.~\eqref{eq:ampl:anti:ll} is the 
four-dimensional result, cf.\ eq.~\eqref{eq:ampl:born:ll:4d}. The second term is 
an evanescent term of $\mathcal{O}(\epsilon)$ and a consequence of the 
prescription~\eqref{eq:completeness:anti}. Note that the evanescent term depends 
directly on the momentum $k^\mu$. For these contributions to be consistently 
generated at all perturbative orders, and to have any chance for their eventual 
cancellation in the finite result, spinor chains must always be combined in a 
fixed order using the same rational expression of unity. This is similar to 
other procedures using anticommuting $\gamma_5$, which require choosing a 
reading point for gamma traces resulting from spin sums and virtual fermion 
loops.

Then, we can write eq.~\eqref{eq:ampl:anti:ll}, as well as the remaining  
helicity amplitudes, using $k^\mu = p_1^\mu$, 
\begin{align}
    \label{eq:ampl:born:mm:final}
    A^{(0),\textrm{ac}}_{--} &\sim \langle 1 | \textrm{P}_+ \gamma^\mu  \textrm{P}_- | 3] 
        \langle 2 | \textrm{P}_- \gamma_\mu  \textrm{P}_+ | 4 ]  
        = -2 \langle 1 2 \rangle [34] + 2 \epsilon\frac{\langle 1 2 \rangle 
        [13][24]}{[12]} \, , \\
    \label{eq:ampl:born:mp:final}
    A^{(0),\textrm{ac}}_{-+} &\sim \langle 1 | \textrm{P}_+ \gamma^\mu  \textrm{P}_- | 3] 
        [ 2 | \textrm{P}_- \gamma_\mu  \textrm{P}_+ | 4 \rangle 
        = +2\langle 14 \rangle [23] - 2\epsilon \frac{\langle 13 \rangle 
        \langle 24\rangle [23]}{\langle 23 \rangle} \, , \\
    \label{eq:ampl:born:pm:final}
    A^{(0),\textrm{ac}}_{+-} &\sim  [ 1 | \textrm{P}_- \gamma^\mu  \textrm{P}_+ | 3\rangle 
        \langle 2 | \textrm{P}_+ \gamma_\mu  \textrm{P}_- | 4 ] 
        = +2 \langle 23 \rangle [14] - 2 \epsilon \frac{\langle 23 \rangle 
        [13][24]}{[23]} \, , \\
    \label{eq:ampl:born:pp:final}
    A^{(0),\textrm{ac}}_{++} &\sim  [ 1 | \textrm{P}_- \gamma^\mu  \textrm{P}_+ | 3\rangle 
        [ 2 | \textrm{P}_- \gamma_\mu  \textrm{P}_+ | 4 \rangle 
        = -2 \langle 34\rangle [12] + 2\epsilon \frac{\langle 13\rangle 
        \langle 24\rangle [12]}{\langle 12\rangle} \, .
\end{align}
To arrive at eqs.~(\ref{eq:ampl:born:mp:final}~-~\ref{eq:ampl:born:pp:final}), we 
inserted  unity expressed as $1 = \langle 3 | \textrm{P}_+ \textrm{P}_+ | 2\rangle/\langle 3 | 
\textrm{P}_+ \textrm{P}_+ | 2\rangle$, $1 = [3 | \textrm{P}_- \textrm{P}_- | 2]/[3 | \textrm{P}_- \textrm{P}_- | 2]$, and $1 = [ 3 | 
\textrm{P}_- \slashed{1} \textrm{P}_+ |2\rangle/[3 | \textrm{P}_- \slashed{1} \textrm{P}_+ | 2\rangle$ 
respectively. Note that, for some helicity configurations, no auxiliary momentum 
$k^\mu$ is needed.

This procedure can be trivially extended to multi-loop computations. However, at 
loop level, the two spinor chains might not be fully contracted with each other, 
but may be contracted with tensor loop-integrals. To reduce the one-loop tensor integrals in this computation, we split the loop momentum
\begin{align}
    \label{eq:loop:par_trans}
    l^\mu = l_\parallel^\mu + l_\perp^\mu \, ,
\end{align}
into a parallel component $l_\parallel^\mu$, that is part of the span of momenta $\{q_i\}$ that are present in the propagators, $d_i \equiv (l + q_i)^2 - m_i^2$, of the integral 
family, and a transversal component $l_\perp^\mu$, with $l_\perp \cdot q_i = 0$.
In general, the $q_i$ are linear combinations of four-dimensional external momenta $p_i$.
The former can be expressed as sum of inverse propagators multiplying 
momenta $q_i^\mu$, most conveniently with the help of the van~Neerven-Vermaseren 
basis~\cite{vanNeerven:1983vr},
\begin{align}
    \label{eq:vNV_basis}
    l_\parallel^\mu = \sum_i (l_\parallel \cdot q_i) \, v_i^\mu = \sum_i (l \cdot q_i) \, v_i^\mu \, ,
\end{align}
where $v_i^\mu$ are the van~Neerven-Vermaseren basis vectors, that are linear combinations of $q_i^\mu$ with rational functions in kinematic invariants as coefficients. 
By definition of the set $\{q_i\}$, all scalar products $(l \cdot q_i)$ in eq.~\eqref{eq:vNV_basis} can be expressed in terms of inverse propagators.
All remaining open tensor indices are related to $l_\perp$, and as an example we consider an integral with numerator $l_\perp^\mu l_\perp^\nu$. By construction, inverse propagators ${d_i}$ only depend on $l_\perp^2$ and not its direction.
Consequently, we can make the ansatz
\begin{align}
    \label{eq:mastrolia}
    \int \textrm{d}^{d_\perp}l_\perp \frac{l_\perp^\mu l_\perp^\nu}{\prod d_i} = C g_\perp^{\mu\nu} \quad \Rightarrow \quad C = \frac{1}{d_\perp} \int \textrm{d}^{d_\perp}l_\perp \frac{l_\perp^2}{\prod d_i} \, ,
\end{align}
where $d_\perp$ is the dimension and $g^{\mu\nu}_\perp$ is the metric tensor of the transverse space.
To find $C$, we contracted both sides with $g^\perp_{\mu\nu}$.
This procedure can be easily extended to any loop order and any number of ranks~\cite{Mastrolia:2016dhn}.
This might require more complex ans\"atze with multiple possible combinations of $g^\perp_{\mu\nu}$, and finding the scalar integral coefficients by solving a linear system of equations whose coefficients are polynomials in $d_\perp$, but the overall principle is identical.
This decomposition introduces combinatorical complexity for higher tensor ranks, that can, however, be reduced by exploiting symmetry properties of the tensors, see e.g.\ refs.~\cite{Goode:2024mci, vonManteuffel:2025swv}.

In order to replace the remaining $l_\perp^2$ on the right-hand side of eq.~\eqref{eq:mastrolia}, we can expand
$g_\perp^{\mu\nu}$ in terms 
of $d$-dimensional objects,
\begin{align}
    \label{eq:gperp}
    g^{\mu\nu}_\perp = g^{\mu\nu} - \sum_{i = 1}^{d_\parallel} p_i^\mu v_i^\nu \, ,
\end{align}
where $v_{i=1\dots4}^\mu$ are the van~Neerven-Vermaseren basis vectors of the 
span of $\{q_i\}$ and $d_\parallel$ the dimension of the latter. 
This allows us to write
\begin{align}
    l_\perp^2 = l_\mu l_\nu g^{\mu\nu}_\perp = l^2 - \sum_{i = 1}^{d_\parallel} (q_i \cdot l) (v_i \cdot l) \, , 
\end{align}
such that, in a second step, $l_\perp^2$ can be expressed in terms of inverse propagators.

Let us comment on another issue with respect to computational complexity.
For the pentagon topologies of non-factorizable VBF, momenta $q_i$ span the full four-dimensional space-time, $d_\parallel = 4$.
Hence, $d_\perp = d - d_\parallel \sim \epsilon$, and we can see from eq.~\eqref{eq:mastrolia} that reducing tensor integrals by averaging over transversal components can lead to spurious $1/\epsilon$ poles.
These poles always multiply instances of $g_\perp^{\mu\nu}$, that also vanishes for $\epsilon \rightarrow 0$.
However, immediate application of eq.~\eqref{eq:gperp} can lead to
large intermediate expressions, that 
have to cancel as only $\mathcal{O}(\epsilon)$ contributions of 
eq.~\eqref{eq:gperp} can contribute. We, therefore, find it useful to keep 
$g_\perp^{\mu\nu}$ as long as possible, and use the expansion 
eq.~\eqref{eq:gperp} only once necessary.

We stress that, if we combine spinor chains first, much of the tensor integral 
reduction can be avoided using Dirac gamma algebra. For example, let us consider 
a contribution to the non-factorisable Feynman 
diagram depicted in figure~\ref{fig:1loop}~(b)~(top-left),
\begin{align}
    \label{eq:tensor}
    \langle 1| \textrm{P}_+ \gamma^\rho \gamma^\mu \gamma^\sigma \textrm{P}_- |3]  
    \langle 2| \textrm{P}_+ \gamma_\rho \gamma^\nu \gamma_\sigma \textrm{P}_- |4] 
    \times\int \textrm{d}^4l \, \frac{l_\mu l_\nu}{d_1 d_2 d_3 d_4 d_5} \, ,
\end{align}
where $d_{i = 1 \dots 5}$ are inverse propagators of this diagram, whose 
explicit forms are not relevant for the following arguments. We can continue 
reducing the tensor integral in eq.~\eqref{eq:tensor}, which requires a rank-2 tensor decomposition.
Alternatively, we can 
first combine spinor chains
\begin{align}
    \label{eq:tensorred:step1}
    \langle 1| \textrm{P}_+ \gamma^\rho \slashed{l} \gamma^\sigma \textrm{P}_- |3]  
    \langle 2| \textrm{P}_+ \gamma_\rho \slashed{l} \gamma_\sigma \textrm{P}_- |4] 
        = \frac{\langle 1 | \textrm{P}_+ \gamma^\rho \slashed{l} \gamma^\sigma 
            \slashed{3} \slashed{1} \slashed{2} \gamma_\rho \slashed{l} 
            \gamma_\sigma \textrm{P}_- | 4]}{\langle 3 | \slashed{1} | 2 ]} \, ,
\end{align}
and anticommute the two instances of $\slashed{l}$ until they are next to 
each other and the relation $\slashed{l} \slashed{l} = l^2$ can be used.
This produces terms in which the loop momentum enters only either through $l^2$ or through $\slashed{l}$ times scalar products $l \cdot p_i$.
The scalar products involving $l$ can be written as linear combinations of inverse propagators $d_i$ and invariants, and we find
\begin{align}
    \label{eq:tensorred:step2}
    \langle 1 | \textrm{P}_+ \gamma^\rho \slashed{l} \gamma^\sigma \slashed{3} 
        \slashed{1} \slashed{2} \gamma_\rho \slashed{l} \gamma_\sigma \textrm{P}_- | 4] 
    = l_\mu \bigg( a_0^\mu + \sum_{i=1}^5 a_i^\mu d_i \bigg) 
        + \bigg( b_0 + \sum_{i=1}^5 b_i \, d_i \bigg) \, ,
\end{align}
where $a_{i=0\dots5}^\mu$ and $b_{i=0\dots5}$ are coefficients that do not 
depend on the loop momentum $l$. The first step of this procedure, combining 
slashed loop momenta in a spinor chain, works in the same way for multi-loop 
integrals as well. It can be applied whenever we have more than one instance of 
the same loop momentum contracted into a spinor chain. For integral topologies 
where the parallel space coincides with the space spanned by the external 
momenta and polarisation vectors, we can also generalize the second step, and 
express all scalar products, that we might pick up by commuting the slashed loop 
momentum, through inverse propagators. We conclude that, in such cases, tensor 
decompositions of integrals with a rank higher than the loop order are never 
required. In particular, this is trivially the case for integral topologies with 
more than four legs. In the current one-loop computation, the reduction of the 
remaining rank-one integrals is trivial, given that integrals with a single 
transversal rank vanish identically,
\begin{align}
    \int \textrm{d}^{d_\perp} l_\perp \, \frac{ l_\perp^{\mu}}{d_1 \dots d_5} = 0 \, .
\end{align}
This follows directly from the fact that the only tensor that can be used in the ansatz~\eqref{eq:mastrolia} is the rank-$2$ tensor $g_\perp^{\mu\nu}$.

The set of scalar integrals is then reduced to a minimal set of master integrals 
using IBP methods. The complexity of the reduction can depend on the chosen 
method. For example, factorisable diagrams have a simpler triangle subtopology, 
and contributing integrals can be up to rank two due to the two fermion 
propagators. The projector method, discussed in section~\ref{sec:projectors}, 
which uses all four external momenta, might introduce unreduced triangles with 
up to two inverse propagators if the projection directions lie outside the 
triangle topology. In contrast, a tensor reduction of the integrals preserves the 
fundamental topology and prevents the appearance of higher-rank numerators, 
resulting in a simpler integration-by-parts reduction.
The spinor-helicity methods presented here help to reduce the rank of the tensor decomposition of the loop integrals.

Finally, we note that this procedure is equivalent to the naive approach of not 
adding any additional projectors, using anticommuting $\gamma_5$, and simply 
prescribe that
\begin{align}
\begin{aligned}[b]
    \textrm{P}_-| i\rangle &= 0 \, , \quad \textrm{P}_+| i ] = 0 \, , \\
    \langle i | \textrm{P}_- &= 0 \, , \quad [i|\textrm{P}_+ = 0 \, ,
\end{aligned}
\end{align}
still holds. What we learned from these considerations, is the similarity of the expression of unity and combination of spinor chains with the fixing of a reading point when evaluating Dirac traces in dimensional regularization with anticommuting $\gamma_5$. In the following section, we repeat this discussion for the 
BMHV $\gamma_5$ scheme.


\subsection{Breitenlohner-Maison / 't~Hooft-Veltman \texorpdfstring{\boldmath $\gamma_5$}{gamma 5}}
\label{sec:helamp:thooft-veltman}

We repeat the computation of helicity amplitudes~(\ref{eq:ampl:born:mm:final}~-~\ref{eq:ampl:born:pp:final}) using a 
BMHV $\gamma_5$ scheme.
An expression of unity similar to eq.~\eqref{eq:unity} for the BMHV $\gamma_5$ scheme has been introduced in the \texttt{Spinney Form} library~\cite{Cullen:2010jv} to turn spinor chains into Dirac traces. Here, however, our goal is to combine spinor chains.   

In this scheme,
\begin{align}
    \label{eq:BMHV:gamma}
    \gamma^\mu = \bar{\gamma}^\mu + \hat{\gamma}^\mu \, , \quad g^{\mu\nu} = 
        \bar{g}^{\mu\nu} + \hat{g}^{\mu\nu} \, ,
\end{align}
where $\bar\gamma^\mu$ are treated as the usual $4$-dimensional gamma matrices 
and $\hat\gamma^\mu$ are evanescent components with 
\begin{align}
\label{eq:BMHV:gamma5:commutators}
\begin{aligned}[b]
    \{\bar{\gamma}^\mu, \bar{\gamma}^\nu\} &= 2\bar{g}^{\mu\nu} \, , &    
    \{\hat{\gamma}^\mu, \hat{\gamma}^\nu\} &= 2\hat{g}^{\mu\nu} \, , \\
    \{\gamma_5, \bar{\gamma}^\mu\} &= 0 \, , & 
     [\gamma_5, \hat{\gamma}^\mu]  &= 0 \, ,
\end{aligned}
\end{align}
where $\bar{\gamma}^\mu \bar{\gamma}_\mu = 4$, $\hat{\gamma}^\mu 
\hat{\gamma}_\mu = d - 4$, $\bar{g}^{\mu\nu} \bar{g}_{\mu\nu} = 4$, and 
$\hat{g}^{\mu\nu} \hat{g}_{\mu\nu} = d - 4$. To preserve $\{\gamma^\mu, 
\gamma^\nu\} = 2g^{\mu\nu}$, we also need
\begin{align}
\label{eq:BMHV:mix}
    \{\bar{\gamma}^\mu, \hat{\gamma}^\nu\} = 0 \, .
\end{align}
Since $\gamma_5$ is treated as the usual $4$-dimensional object, the action of 
projectors on helicity spinors is well defined,
\begin{align}
\label{ep:proj:BMHV}
\begin{aligned}[b]
    \textrm{P}_- | i] &= |i] \, , & \quad  \textrm{P}_+ | i \rangle &= | i \rangle  \, , \\
    \textrm{P}_+ | i] &= 0 \, , & \quad  \textrm{P}_- | i \rangle &= 0 \, . 
\end{aligned}
\end{align}

We start again from the amplitude~\eqref{eq:ampl:born:ddim}.
In this $\gamma_5$-scheme, we use unity expressed as
\begin{align}
    \label{eq:unity:BMHV}
    1 = \frac{\langle 3 | \slashed{\bar{k}} | 2]}{\langle 3 | 
    \slashed{\bar{k}} | 2]} \, ,
\end{align}
where we defined $\slashed{\bar{k}} \equiv k_\mu \bar{\gamma}^\mu$. Inserted 
into the amplitude~\eqref{eq:ampl:born:ddim}, we find
\begin{align}
    \label{eq:ampl:BMHV}
        A^{(0),\textrm{BMHV}}_{--} &\sim \frac{
        \langle 1 | \gamma^\mu |3] \langle 3|\slashed{\bar{k}}|2] 
        \langle 2 | \gamma_\mu |4]}{\langle 3|\slashed{\bar{k}}|2]} \,,
\end{align}
where we removed projectors next to external spinors according to eq.~\eqref{ep:proj:BMHV}.
The superscript BMHV indicates a $d$-dimensional amplitude computed in the 
BMHV $\gamma_5$-scheme. In this scheme,
\begin{align}
    \label{eq:completeness:BMHV:bare}
    \slashed{\bar{p}}_i = |i\rangle[i| + |i]\langle i| \, .
\end{align}
We can not use this directly in eq.~\eqref{eq:ampl:BMHV} to combine spinor 
chains, because contributions of the second helicity are missing, and there is 
no mechanism that would cancel them as long as $d$-dimensional objects are 
present. However, in contrast to the anticommuting scheme, actions of projectors 
eq.~\eqref{ep:proj:BMHV} are well-defined in this scheme, and we write 
eq.~\eqref{eq:completeness:BMHV:bare} as
\begin{align}
    \label{eq:completeness:BMHV}
    |i]\langle i| = \textrm{P}_- \slashed{\bar{i}} \textrm{P}_+ \, , \quad 
    |i\rangle [i| = \textrm{P}_+ \slashed{\bar{i}} \textrm{P}_- \, .
\end{align}
Inserted into eq.~\eqref{eq:ampl:BMHV}, we find
\begin{align}
        A^{(0),\textrm{BMHV}}_{--} &\sim \frac{  \langle 1|  \gamma^\mu \textrm{P}_- 
            \slashed{\bar{3}} \textrm{P}_+ \slashed{\bar{k}} \textrm{P}_- \slashed{\bar{2}} 
            \textrm{P}_+ \gamma_\mu | 4]}{\langle 3 | \slashed{\bar{k}} | 2]} 
        = -2 \langle 1 2 \rangle [34] \, ,
\end{align}
where we used the splitting of $\gamma^\mu$~\eqref{eq:BMHV:gamma}, and (anti-)commutators~(\ref{eq:BMHV:gamma5:commutators},~\ref{eq:BMHV:mix}) to bring $\gamma$-matrices with contracted indices together, as well as Schouten identities~\eqref{eq:schouten} to simplify the 
result once everything is four-dimensional. The result is identical to the four-dimensional result~\eqref{eq:ampl:born:ll:4d}.
Furthermore, it is independent of $k^\mu$, since 
eq.~\eqref{eq:completeness:BMHV} is exact for BMHV 
$\gamma_5$ provided $k^\mu$ is a physical, massless momentum.

We continue with the one-loop computation. As an instructive example, we 
consider a spinor structure that factorizes from a one-loop integral in amplitude 
$A^{(1),\textrm{BMHV}}_{--}$ originating from the diagram depicted in figure~\ref{fig:1loop}~(a)~(top),
\begin{align}
    g_{q_1}^- g_{q_2}^+ \, \langle 1 | \gamma^\mu|3] \langle 2 | 
        \gamma^\nu \gamma^\rho \gamma_\mu \textrm{P}_+ \gamma_\rho \gamma_\nu | 4] 
        =  -16 \epsilon \,   g_{q_1}^- g_{q_2}^+  \, \langle 1 2 \rangle [34] \, .
\end{align}
This contribution contains ``wrong'' couplings for quarks with the considered helicities and, therefore, should not contribute to the amplitude $A^{(1),\textrm{BMHV}}_{--}$ in four dimensions.
Nevertheless, this, and similar 
contributions might start to contribute at $\mathcal{O}(\epsilon^0)$ if they are multiplied with 
poles from loop integrals. This is not uncommon for computations in the BMHV 
scheme and counterterms need to be added. Since helicity is conserved at all 
orders and unphysical contributions depend on different couplings, these terms 
can be consistently set to zero, and we choose to do so. Nevertheless, 
counterterms are also required for physical helicities.

In this scheme, couplings $g_q^\pm$ have to be considered bare 
$\bar{g}_{q}^\pm$. We obtain the correct counterterms at NLO QCD by 
replacing~\cite{Larin:1993tq}
\begin{align}
    \bar{g}_{q}^- &\rightarrow g_{q}^- \bigg(1 - \frac{\alpha_s}{2\pi} \, C_f 
        + \mathcal{O}\big(\alpha_s^2\big)\!\bigg) \, , \nonumber\\
    \bar{g}_{q}^+ &\rightarrow g_{q}^+ \bigg(1 + \frac{\alpha_s}{2\pi} \, C_f 
        + \mathcal{O}\big(\alpha_s^2\big)\!\bigg) \, ,
    \label{eq:coupling:counter}
\end{align}
in the Born amplitude. Note that, if we followed ref.~\cite{Larin:1993tq} 
verbatim, e.g.\ the replacement for $\bar{g}_{q}^-$ in 
eq.~\eqref{eq:coupling:counter} would also contain terms proportional to 
$g_{q}^+$, mixing couplings from different helicities. Here we have set them to 
zero, consistent with the computation of the one-loop helicity amplitudes. 
It would be interesting to apply the spinor method discussed here in the context of a rigorous and general renormalisation program for the BMHV scheme, see refs.~\cite{OlgosoRuiz:2024dzq,Ebert:2024xpy,Kuhler:2025znv, vonManteuffel:2025swv} for recent progress in this direction.

We further note that, also in case of the BMHV scheme, we can avoid tensor 
integral reduction. As example, we consider the same loop 
integral~\eqref{eq:tensor} for the BMHV $\gamma_5$ scheme as in the previous  
section for anticommuting $\gamma_5$,
\begin{align}
    \langle 1 |  \gamma^\rho \slashed{l} \gamma^\sigma  | 3]  \langle 2|  
        \gamma_\rho \slashed{l} \gamma_\sigma  | 4] 
    = \frac{\langle 1 |  \gamma^\rho \slashed{l} \gamma^\sigma \textrm{P}_- 
        \slashed{\bar{3}} \textrm{P}_+ \slashed{\bar{1}} \textrm{P}_- \slashed{\bar{2}} 
        \textrm{P}_+ \gamma_\rho \slashed{l} \gamma_\sigma  | 4]}{\langle 3 | 
        \slashed{\bar{1}} | 2 ]} \, . 
\end{align}
To proceed, we have to split the loop momentum in a four- and a 
$(-2\epsilon)$-dimensional component,
\begin{align}
    \label{eq:dirac:phys_trans}
    l^\mu = \bar{l}^\mu + \hat{l}^\mu \, .
\end{align}
Since contractions of the loop momentum are all within the same spinor chain, we can replace the loop momentum according to eq.~\eqref{eq:dirac:phys_trans} and expand.
Termwise, we bring repeated instances of momenta $\bar{l}$ or $\hat{l}$ together using commutators and anticommutators~\eqref{eq:BMHV:gamma5:commutators}, respectively, and replace $\slashed{\bar{l}}\slashed{\bar{l}} = \bar{l}^2$ and $\slashed{\hat{l}}\slashed{\hat{l}} = \hat{l}^2$. Four structures might remain as numerators of the loop integrals:
\begin{align}
    \label{eq:BMHV:structures}
    \bar{l}^2\, , \ \hat{l}^2 \, , \ \bar{l}^\mu \hat{l}^\nu \, , \ 
        \bar{l}^\nu \hat{l}^\mu \, .
\end{align}
For a process, like VBF, the splitting of the $d$-dimensional space in the BMHV scheme for Dirac algebra, coincides with the splitting into parallel and transversal space in the loop-integrals, cf.\ eq.~\eqref{eq:loop:par_trans}.
We can, therefore, treat the numerators~\eqref{eq:BMHV:structures} in the same way as discussed for the anticommuting $\gamma_5$-scheme.\footnote{This remains true for processes with larger 
numbers of external legs. For processes with fewer external legs, the last two 
structures in eq.~\eqref{eq:BMHV:structures} still vanish as they are part of a 
subspace of the transversal space. One of $\bar{l}^2$ or $\hat{l}^2$ can be 
easily expressed through the other and inverse propagators. However, one 
structure remains that needs to be reduced in the usual way. At this point, 
benefits of this approach largely vanish.}

As final step, scalar integrals are reduced to a set of master integrals using 
IBP reduction. Similar to the case of the anticommuting $\gamma_5$ scheme, the IBP reduction 
is simpler as in case of the projector method. It remains to compute the set of 
master integrals, which we discuss in detail in the next section.


\section{Master integrals}
\label{sec:master-integrals}

In the previous sections, we discussed different methods to compute the 
$d$-dimensional amplitudes and to reduce tensor integrals to scalar integrals. 
The latter can then be reduced by integration by 
parts~\cite{Chetyrkin:1981qh, Laporta:2000dsw} onto a smaller number of master 
integrals.
Here, we use the public program \texttt{Reduze~2}~\cite {vonManteuffel:2012np}.

It now remains to evaluate these scalar integrals to higher orders in 
$\epsilon$. While it is possible to express the finite part of the amplitudes in 
terms of known integrals (see e.g.\ ref.~\cite{tHooft:1978jhc}), for higher orders 
this is not the case and genuine five-point integrals are required.

\begin{figure}[t]
    \centering
    \includegraphics[height=110pt]{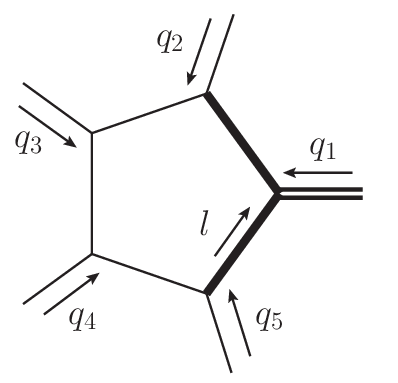} \\[-10pt]
    \caption{One-loop pentagon integral employed in this computation. Thick lines 
    correspond to internal particles with mass square $m^2$, the double line to 
    an external particle with mass square $\bar{s}_1$, and all other lines are 
    massless.}
    \label{fig:pentagon:masterintegrals}
\end{figure}

All occurring integrals can be mapped onto the pentagon family, see 
figure~\ref{fig:pentagon:masterintegrals},
\begin{align}
    \label{eq:master:topo}
    I_{\nu_1 \nu_2 \nu_3 \nu_4 \nu_5}(\vec{x}) 
    \equiv e^{\epsilon \gamma_E} \mu_0^{2\epsilon} \int 
    \frac{\mathrm{d}^d l}{\mathrm{i} \pi^{d/2}} \frac{1}{d_1^{\nu_1} 
    d_2^{\nu_2} d_3^{\nu_3} d_4^{\nu_4} d_5^{\nu_5}} \, ,
\end{align}
with
\begin{align}
     d_1 \equiv l^2-m^2 \, , \quad 
     d_2 \equiv (l + q_1)^2 - m^2 \, , \quad 
     d_n\equiv \bigg(l+\sum_{i=1}^{n-1} q_i\bigg)^2 \quad 
        \mathrm{for}\ n\in\{3,4,5\} \, , 
\end{align}
where $\mu_0$ is an arbitrary scale introduced such that the integrals have 
integer mass dimension also away from $d=4$, and $\gamma_E$ is the 
Euler-Mascheroni constant. These integrals are described in terms of a vector of 
seven independent kinematic invariants,
\begin{align}
    \vec{x}\equiv(m^2, \bar{s}_1, \bar{s}_{12}, \bar{s}_{23}, \bar{s}_{34}, \bar{s}_{45}, \bar{s}_{15}) \, ,
\end{align}
where $m^2$ is the internal mass square, $\bar{s}_1\equiv q_1^2$, and $\bar{s}_{ij} \equiv 
(q_i + q_j)^2$ (we use the overbar to distinguish these Mandelstam invariants from
$s_{ij}$ defined in terms of vectors $p_i$). 
In the complex-mass scheme \cite{Denner:1999gp}, $m^2$ becomes a 
complex quantity while all other entries of $\vec{x}$ remain real. Nevertheless, 
to perform analytic continuations, in intermediate steps we equip all kinematic 
invariants with non-zero imaginary parts.
As the mapping from the kinematics of figure~\ref{fig:born} to figure~\ref{fig:pentagon:masterintegrals}, we choose
\begin{align}
    p_1 \rightarrow -q_5 \, , \quad p_2 \rightarrow - q_4 \, , \quad
    p_3 \rightarrow -q_3 \, , \quad p_4 \rightarrow - q_2 \, , \quad
    p_5 \rightarrow -q_1 \, .
\end{align}

The master integral basis for the family \eqref{eq:master:topo} contains $18$ 
integrals, and we compute them using $\epsilon$-factorised differential 
equations \cite{Kotikov:1990kg, Henn:2013pwa}. Using integrand analysis (see e.g. 
ref.~\cite{Henn:2020lye}) in a Baikov representation \cite{Baikov:1996iu}, we 
find the $\epsilon$-factorising basis
\begin{align}
    \begin{aligned}[b]
    I_1    &= \epsilon   \, I_{20000} \, , & 
    I_2    &= \epsilon   \, r_4 I_{21000} \, , \\
    I_3    &= \epsilon^2 \, (\bar{s}_1  - \bar{s}_{12}) \, I_{11100} \, , & 
    I_4    &= \epsilon^2 \, r_3\, I_{11010} \, , \\
    I_5    &= \epsilon^2 \, (\bar{s}_{12} - \bar{s}_{45}) \, I_{10110} \, , & 
    I_6    &= \epsilon^2 \, \bar{s}_{23} \, I_{01110} \, , \\
    I_7    &= \epsilon^2 \, (\bar{s}_1  - \bar{s}_{15}) \, I_{11001} \, , & 
    I_8    &= \epsilon^2 \, (\bar{s}_{12} - \bar{s}_{34}) \, I_{10101} \, , \\
    I_9    &= \epsilon^2 \, (\bar{s}_{15} - \bar{s}_{34}) \, I_{01101} \, , & 
    I_{10} &= \epsilon^2 \, \bar{s}_{45}\, I_{10011}\,, \\
    I_{11} &= \epsilon^2 \, (\bar{s}_{15} - \bar{s}_{23})\, I_{01011}\,, & 
    I_{12} &= \epsilon^2 \, \bar{s}_{34}\, I_{00111}\,, \\
    I_{13} &= \epsilon^2 \, (m^2  \bar{s}_{12} + m^2  \bar{s}_{23} - \bar{s}_{12} \bar{s}_{23} 
        - m^2  \bar{s}_{45})\, I_{11110}\,, & \quad  
    I_{14} &= \epsilon^2\, r_2\, I_{11101}\,, \\
    I_{15} &= \epsilon^2\, (m^2  \bar{s}_{15} - m^2  \bar{s}_{23} + m^2  \bar{s}_{45} 
        - \bar{s}_{15} \bar{s}_{45})\, I_{11011}\,, & I_{16} &= \epsilon^2\, 
        \bar{s}_{34}\, (m^2  - \bar{s}_{45})\, I_{10111}\,, \\
    I_{17} &= \epsilon^2\, (m^2  - \bar{s}_{23})\, \bar{s}_{34}\, I_{01111} \, , \\
    I_{18} &= \epsilon^2 (8 \, r_1)^{-1}\,(Q_0\, I_{11111} 
        + Q_1 \, I_{01111} + Q_2\, I_{10111} + Q_3\, I_{11011} 
        + Q_4\, I_{11101} + Q_5\, I_{11110})\,. \hspace{-300pt}
    \end{aligned} \nonumber\\
\label{eq:MI:canonical-basis}
\end{align}
Here, $Q_i$ are polynomials in kinematic invariants whose explicit forms can be 
found in appendix~\ref{app:basisdefpolys}. Moreover, we have defined four square 
roots $r_{i=1\dots4}$, which satisfy
\begin{align}
    \begin{aligned}[b]
    r_1^2&= 16\,\bar{\Delta}_5\,,\\
    r_2^2 &= m^4 \bar{s}_{12}^2 + 2\, m^4 \bar{s}_{12} \bar{s}_{15} - 2\, m^2 \bar{s}_{12}^2 \bar{s}_{15} 
        + m^4 \bar{s}_{15}^2  - 2\, m^2 \bar{s}_{12} \bar{s}_{15}^2 + \bar{s}_{12}^2 \bar{s}_{15}^2 \\
        &\hspace{10pt}- 4\, m^4 \bar{s}_1 \bar{s}_{34} + 2\, m^2 \bar{s}_1 \bar{s}_{12} \bar{s}_{34} 
        + 2\, m^2 \bar{s}_1 \bar{s}_{15} \bar{s}_{34} + 4\, m^2 \bar{s}_{12} \bar{s}_{15} \bar{s}_{34} \\
        &\hspace{10pt}- 2\, \bar{s}_1 \bar{s}_{12} \bar{s}_{15} \bar{s}_{34} - 4\, m^2 \bar{s}_1 
        \bar{s}_{34}^2 + \bar{s}_1^2 \bar{s}_{34}^2 \, , \\
    r_3^2&=\bar{s}_1^2 + \bar{s}_{23}^2 + \bar{s}_{45}^2 - 2\, \bar{s}_1 \bar{s}_{23} - 2\, \bar{s}_1 \bar{s}_{45} 
        - 2\, \bar{s}_{23} \bar{s}_{45}\,, \\
    r_4^2 &= -\bar{s}_1\,(4\, m^2 - \bar{s}_1) \, .
    \end{aligned}
    \label{eq:square_roots}
\end{align}
Here,
\begin{align}
    \bar{\Delta}_5 & \equiv 
        \begin{vmatrix}
            q_1 \cdot q_1 & \dots & q_1 \cdot q_4 \\
            \vdots & \ddots & \vdots \\
            q_4 \cdot q_1 & \dots & q_4 \cdot q_4
        \end{vmatrix} 
\end{align}
is the five-point Gram determinant for the kinematics considered in the master integral computation.
The overall $\epsilon$-dependent prefactor is chosen such that the leading term 
of the integral basis is $\mathcal{O}(\epsilon^0)$. For convenience, we also 
provide the basis~\eqref{eq:MI:canonical-basis} as an ancillary file in \texttt{Reduze~2} format, where we omit the multi-valued 
square-root prefactors.

The canonical differential equation for the basis $\vec{I} \equiv (I_1, \dots, 
I_{18})^T$ takes on the form
\begin{align}
\label{eq:MI:MIdeq}
    \mathrm{d}\vec I=\epsilon\sum_k \mathrm{d}\log(W_k)\,A_k\,\vec{I}\,,
\end{align}
where the letters $W_k$ are algebraic functions of kinematic invariants and 
$A_k$ are $\mathbb{Q}$-valued matrices. In principle, we could derive the differential equation~\eqref{eq:MI:MIdeq} by direct computation, however this would require manual 
identification of the $\mathrm{d}\log$-forms appearing in it and consequently 
tedious manipulations of partial derivative matrices. Therefore, we proceed by 
taking eq.~\eqref{eq:MI:MIdeq} as an ansatz and fitting the matrices $A_k$ from 
numerical samples (see e.g.\ \cite{Abreu:2018rcw}). This requires the knowledge 
of the letters $W_k$, which we determine by using a combination of the public 
code \texttt{Baikovletter}~\cite{Jiang:2024eaj}, the one-loop alphabets given in 
\cite{Dlapa:2023cvx}, the algorithm described in \cite{Heller:2019gkq, 
Matijasic:2024too}, and in remaining cases by explicit integration of partial 
derivatives.

Note that the signs of square roots~\eqref{eq:square_roots} are 
independently arbitrary, but integrals with square-root prefactors, $I_2$, 
$I_4$, $I_{14}$, and $I_{18}$, depend on these signs. This allows us to simplify 
the ansatz~\eqref{eq:MI:MIdeq}, as the letters $W_k$ must have the correct 
transformation behaviour under sign changes of square roots $r_i$. As an 
example, consider the differential $\mathrm{d} I_2\propto r_4$ that changes sign 
for $r_4 \rightarrow -r_4$. Since $I_2$ is a bubble integral, its only 
subtopology contribution comes from the tadpole integral $I_1$. Hence, we have
\begin{align}
    \mathrm{d}I_2=\epsilon\,(\omega_1\,I_1+\omega_2\,I_2)\,,
\end{align}
where $\omega_i$ are linear combinations of $\mathrm{d}\log$-forms. Here, 
$\omega_1$ must change sign for $r_4 \rightarrow -r_4$, since $I_1$ is 
independent of the square root $r_4$. Contrarily, $\omega_2$ must be invariant 
as it multiplies $I_2$ that already transforms correctly. We can make this 
manifest with the ansatz
\begin{align}
    \label{eq:MI:deqExample}
    \omega_1=\sum_k a_k\, \mathrm{d}\log\bigg(\frac{q_k+r_4}{q_k-r_4}\bigg)
        \,,\qquad \omega_2=\sum_k b_k\, \mathrm{d}\log(q_k)\,,
\end{align}
where $q_k$ are polynomials in kinematic invariants, and $a_k,\ b_k \in 
\mathbb{Q}$ are coefficients. In particular, the ansatz for $\omega_2$ contains 
$2 \mathrm{d}\log(r_4) = \mathrm{d}\log(-\bar{s}_1(4\,m^2-\bar{s}_1))$.

As a second example, we consider contributions to the differential 
$\mathrm{d}I_4\propto r_3$, that are proportional to $I_2\propto r_4$. Such ones 
can only arise if the corresponding prefactor flips sign if $r_3\to -r_3$ or 
$r_4\to -r_4$. As ansatz we choose
\begin{align}
    \label{eq:MI:deqExample2}
    \mathrm{d}I_4=\epsilon \sum_k a_k\,\mathrm{d}\log\bigg(\frac{q_k+r_3r_4}
        {q_k-r_3 r_4}\bigg) I_2+\textnormal{(terms independent of $I_2$)}\,.
\end{align}
Such ans\"atze for letters constrain the entries where the matrices $A_k$ in 
eq.~\eqref{eq:MI:MIdeq} are allowed to have non-zero entries, and, therefore, 
reduce the number of samples required for the numerical fit. 

The differential equation~\eqref{eq:MI:MIdeq} can be solved analytically, 
allowing the master integrals to be expressed in terms of multiple 
polylogarithms with algebraic arguments. Such a representation has been derived 
up to weight $2$ (i.e.\ up to the 
$\epsilon^2$-coefficient). However, we found this 
representation to be suboptimal for problems with many scales. The analytic 
structure and with it the analytic continuation become far from obvious, as 
polylogarithms may introduce spurious branch cuts whose positions are determined 
by complex expressions of kinematic invariants.

A more robust approach is to instead compute the master integrals directly by solving 
the system of differential equations order-by-order in $\epsilon$ numerically
(see e.g.\ \cite{Czakon:2026tog}). 
This can be further improved, if we eliminate all relationships between the 
$\epsilon$-coefficients of the master integrals, see e.g. 
refs.~\cite{Gehrmann:2018yef, Chicherin:2020oor, Chicherin:2021dyp, 
Abreu:2023rco, Badger:2025ljy}. To this end, we use symbol and coproduct 
manipulations \cite{Goncharov:2005sla, Goncharov:2010jf, 
Duhr:2011zq, Duhr:2012fh} to express all $\epsilon$-coefficients of master 
integrals as graded polynomials in a minimal set of $\mathbb{Q}$-linear 
combinations of these coefficients, which we then treat as new transcendental 
basis functions. For instance, the weight-$2$ component $I_{i,\sigma}^{(2)}$ of 
a master integral $I_i$ in permutation $\sigma$ can be written as
\begin{align}
    \label{eq:MI:fromBasisfunctions}
    I_{i,\sigma}^{(2)}=\sum_{j,k} a^{i,\sigma}_{j,k}\, f_{1,j}\, f_{1,k}
        +\sum_j b^{i,\sigma}_{j}\, f_{2,j}+c^{i,\sigma}\, \zeta_2\,,
\end{align}
where $a^{i,\sigma}_{j,k}$, $b^{i,\sigma}_{j}$, and $c^{i,\sigma}$ are rational 
numbers, $f_{w,j}$ are weight-$w$-basis functions, where $j$ is a numerical 
label, and $\zeta_k\equiv\zeta(k)$ are values of the Riemann zeta function. The 
latter are the only transcendental constants arising in 
eq.~\eqref{eq:MI:fromBasisfunctions} and its analogues up to weight $4$, cf.\ 
ref.~\cite{Abreu:2023rco}. The explicit definitions of our master integrals $I_i$
in terms of $\epsilon$-independent functions $f_{i,j}$ and vice versa are given in an
ancillary file. In principle, we may also give expressions
for the basis functions in terms of multiple polylogarithms,
e.g.
\begin{align}
    f_{1,2} &\equiv I_{12,\textrm{id}}^{(1)} = \log(\bar{s}_{34}) \, .
\end{align}
Here, $\textrm{id}$ denotes the identity permutation.
However, in more complicated cases of such forms, analytic continuation becomes
unclear, as mentioned before.

We consider masters integrals in all possible permutations of external legs 
simultaneously, eliminating redundancies. We find $18$ weight-$1$, $60$ 
weight-$2$, $84$ weight-$3$ and $84$ weight-$4$ functions. Note that not all 
permutations lead to VBF topologies, some also belong to VH with hadronic decay. 
The VBF amplitude alone would only require $9$ weight-$1$, $33$ weight-$2$, $41$ 
weight-$3$ and $41$ weight-$4$ functions. To compute permuted integrals, we also 
require permutations of letters, where we eliminate linear relations between 
them using the integer-relation algorithm of \texttt{Pari/GP}~\cite{Pari:2025}. 
To this end, we also exploit that with our ans\"atze, cf.\ 
eqs.~(\ref{eq:MI:deqExample}~-~\ref{eq:MI:deqExample2}), linear relations may 
only exist between letters whose differential $\mathrm{d}\log$ forms transform 
identically under $r_i \rightarrow-r_i$.

Then, we can define a vector containing all of our transcendental basis functions
\begin{align}
    \vec{f} = (f_{0,1}, f_{1,1}, \dots, f_{1,18}, f_{2,1}, \dots, f_{4,84}) \, ,
\end{align}
where we include $f_{0,1} \equiv 1$ as weight-$0$ function. 
This satisfies an $\epsilon$-independent differential equation,
\begin{align}
    \label{eq:MI:deq}
    \mathrm{d}\,\vec{f}=\sum_{k=1}^{195} \mathrm{d}\log(W_k)\, B_k 
        \vec{f}_\textrm{aug} \, ,
\end{align}
that can be easily derived from eq.~\eqref{eq:MI:MIdeq} using the relations~\eqref{eq:MI:fromBasisfunctions}. 
Here, $\vec{f}_\mathrm{aug}$ is the vector 
$\vec{f}$ augmented with products of lower weight functions, e.g. weight-$3$ products of three weight-$1$ functions. 
The matrices $B_k$ are $\mathbb{Q}$-valued and only couple functions of weight 
$w$ to functions of weight $w-1$. To obtain values for $\vec{f}$ at a 
phase-space point $\vec{x}$, it is sufficient to integrate this differential 
equation along a path $\gamma$ from a boundary point $\vec{x}_0$ to $\vec{x}$,
\begin{align}
    \vec{f}=\vec{f}(x_0)+\int_\gamma\,\,\sum_{k=1}^{195} \mathrm{d}\log(W_k)
        \, B_k \vec{f}_\textrm{aug}\,,
\end{align}
where $\vec{f}(x_0)$ are 
boundary values. The integration is performed numerically, using the 
\texttt{Odeint} library from \texttt{Boost~C++}~\cite{Boost:2026}.

To derive the boundary values $\vec{f}(\vec{x}_0)$, we first compute values for 
the canonical basis~\eqref{eq:MI:canonical-basis} in the identity 
permutation, at the phase-space point
\begin{align}
    \vec{x}_0^\prime\equiv (\mu_0^2,0,0,0,-\mu_0^2,0,0) \, .
\end{align}
This point is well suited because it corresponds to a spurious singularity at 
which many of the $\mathrm{d}\log(W_k)$ are not finite. At the same time, all 
adjacent Mandelstam invariants lie below their corresponding mass thresholds and 
the basis $\vec{I}$ must be finite at $\vec{x}_0^\prime$. This is only possible if 
boundary values of the integrals satisfy linear relations that ensure the 
cancellation of singularities. This fixes all but three boundary constants, and 
we only have to obtain the remaining ones by actual integration. These are the 
tadpole $I_1$, the massive bubble $I_2$ and a massless triangle $I_{12}$, all of 
which can be easily solved at $\vec{x}_0$ using Feynman parameters. We find
\begin{align}
\label{eq:MI:canonicalbasis:boundary}
\begin{aligned}[b]
    I_1(\vec{x}_0^\prime) &= \epsilon\, e^{\epsilon\gamma_E}\Gamma(\epsilon)
        \,, \quad &  
    I_8(\vec{x}_0^\prime)&=I_9(\vec{x}_0^\prime)=-I_1(\vec{x}_0^\prime)
        +I_{12}(\vec{x}_0^\prime) \, , & \\ 
    I_{12}(\vec{x}_0^\prime) &=\epsilon^2\,\frac{-e^{\epsilon\gamma_E}\,
        \Gamma(1+\epsilon)\,\Gamma(-\epsilon)^2}{\Gamma(1-2\epsilon)} \,, \quad & 
    I_{16}(\vec{x}_0^\prime)&=I_{17}(\vec{x}_0^\prime)=I_1(\vec{x}_0^\prime)
        -2I_{12}(\vec{x}_0^\prime) \, , & \\ 
    I_n(\vec{x}_0^\prime) &= 0 \, ,  \quad\textnormal{for} \ 
        n \notin \{1,8,9,12,16,17\} \, . \hspace{-30pt} &
\end{aligned}
\end{align}
Translating this into boundary conditions for $\vec{f}(\vec{x}_0^\prime)$ is 
straightforward by employing the expressions for $f_{i,j}$ in terms of $I_{i,\sigma}$ 
that can be found in auxiliary files. Analogously, we obtain boundary conditions 
for permuted master integrals at appropriate permutations of $\vec{x}_0^\prime$.

The point $\vec{x}_0^\prime$ is far away from the physical region, which is not 
ideal for numerical integration. To improve efficiency, we use 
\texttt{DiffExp}~\cite{Hidding:2020ytt} to transport the boundary conditions to 
a new base-point $\vec{x}_0$. We choose $\vec{x}_0$ to be adjacent to the 
physical region and not inside, with non-vanishing imaginary parts of kinematic 
invariants compatible with Feynman's $i\delta$ prescription. Then, when 
integrating the differential equation along a straight line from $\vec{x}_0$ to 
any physical point $\vec{x}$ with real-valued invariants, the imaginary parts of 
kinematic invariants will only vanish at the very end, effectively circumventing 
singularities.

Although there are no singularities along the integration path, during 
integration, we might cross the branch cut of a square root and we have to make 
sure that the solution remains continuous. This can be done on-the-fly in an algorithmic 
way. As an example, we consider only the root $r_4$ with 
$r_4^2 = -\bar{s}_1 (4 m^2 - \bar{s}_1)$, a base point $\vec{x}_0 = (1-\mathrm{i}/10, -1 + 
\mathrm{i}/10, \dots)$ that we choose to be on the principal branch, and a 
physical point $\vec{x} = (1-\mathrm{i}/10, 1, \dots)$. The integration is along 
a straight path, $\vec{\gamma}(t) = \vec{x}_0 + t \, (\vec{x} - \vec{x}_0)$ for 
$t \in [0,1]$, and we might cross the branch cut if 
\begin{align}
    \label{eq:roots:zeros}
    0 \overset{!}{=} \mathrm{Im}(r_4^2(t)) 
        = -\frac{2}{5} t^2 + \frac{9}{5}t-1 \, .
\end{align}
We solve this polynomial equation numerically and we find a single solution $t_0 
\approx 0.649$ within the interval of interest. To find zeroes also of more complicated polynomials, we use the 
\texttt{FLINT}~library~\cite{flint:2025}.

To understand if we actually cross the branch cut or not, we first check if 
$r_4^2(t_0)$ is on the negative real axis, $\mathrm{Re}(r_4^2(t_0)) \approx 
-1.118 < 0$, which is the case in this example. Second, we check if the path is 
tangent to or actually crosses the negative real axis by comparing signs of 
$\mathrm{Im}(r_4^2(t_0 \pm \delta))$.\footnote{Note that $\delta$ does not need 
to be small. We only have to make sure that $t_0$ is the single zero of the polynomial~\eqref{eq:roots:zeros} between $t_0 - \delta$ and $t_0 + \delta$. In 
this case, since we only cross the real axis once, any value $\delta \in (0, 
1-0.649]$ is fine.} We find that the square root $r_4$ changes from the principal 
branch to its negative, and an analytic continuation is given by
\begin{align}
    r_4(t)=
    \begin{cases}
    \hphantom{-} \sqrt{r_4^2(t)} \, , & \quad 0 \leq t < t_0\,,\\
    - \sqrt{r_4^2(t)} \, , & \quad t_0 \leq t \leq 1\,.
    \end{cases}
\end{align}
In reality, we have to consider all square roots $r_{i=1 \dots 4}$, find the 
zeros of multiple polynomials numerically, and may have to split the integration 
path multiple times, but the algorithmic approach is identical. We note that a 
similar algorithm has been developed in Ref.~\cite{Badger:2025ljy}, however the 
latter relies on multiple evaluations of the square root along the path.

In this way, we can obtain numerical values for the higher-order $\epsilon$ terms of the master integrals in the physical region of phase space.
Both real and complex values can be used for the vector-boson masses.


\section{UV renormalisation and IR subtraction}
\label{sec:uv-ir}

The amplitudes calculated in the previous  sections depend on the bare coupling 
$\alpha_{s,\textrm{b}} = g_{s,\textrm{b}}^2/(4\pi)$, and we renormalise it in 
the standard $\overline{\mathrm{MS}}$ scheme. Since the Born amplitude of this 
process does not depend on $\alpha_{s,\mathrm{b}}$, we only need the replacement
\begin{align}
\label{eq:renorm}
    \alpha_{s,\textrm{b}} \,\mu_0^{2\epsilon}\, S_\epsilon = \alpha_s(\mu)\,
        \mu^{2\epsilon}+ \mathcal{O}(\alpha_s)\,,
\end{align}
to get a UV renormalised result at one-loop order. In eq.~\eqref{eq:renorm}, 
$\mu$ is the renormalisation scale, $\mu_0$ is an arbitrary scale introduced in 
the definition of the integral family~\eqref{eq:master:topo}, and 
$S_\epsilon \equiv (4\pi)^\epsilon\, e^{-\epsilon\gamma_E}$. 
For the sake of brevity, in the following we will set $\mu_0=\mu$.
We write the 
perturbative series in $\alpha_{s}(\mu)$ as
\begin{align}
\label{eq:perturbationseries:renorm}
    \mathcal{A}_\textrm{ren} = \mathcal{A}^{(0)}+\frac{\alpha_s(\mu)}{2\pi}
        \mathcal{A}^{(1)}_\mathrm{ren} + \mathcal{O}(\alpha_s^2) \, .
\end{align}

Infrared poles of the renormalised amplitude factorise, and we define the finite remainder as
\begin{align}
A^{(1),\textrm{f}/\textrm{nf}}_{\lambda,\textrm{fin}} 
        = A^{(1),\textrm{f}/\textrm{nf}}_{\lambda,\textrm{ren}}
        -\textrm{I}^{\textrm{f}/\textrm{nf}}_{1} \, A^{(0)}_{\lambda} 
        \, ,
\end{align}
where we subtract the predicted IR poles~\cite{Giele:1991vf,Kunszt:1994np,Catani:1996vz,Catani:1996mi}
using
\begin{align}
    \textrm{I}_1^\textrm{f} &= -\frac{2}{\epsilon^2}-\frac{1}{\epsilon}
        (3-f_{1,9}-f_{1,10}) \nonumber\\
        &= -\frac{2}{\epsilon^2}-\frac{1}{\epsilon} \bigg(
        3 - \log\bigg(\frac{-s_{13}}{\mu^2}\bigg) - \log\bigg(\frac{-s_{24}}{\mu^2}\bigg)\!\bigg)\,,
    \label{eq:poles:catani:f}\\
    \textrm{I}_1^\textrm{nf} &= \frac{1}{\epsilon}
        (-f_{1,2}+f_{1,3}-f_{1,4}+f_{1,11}) \nonumber\\
        &= \frac{1}{\epsilon} \bigg( -\log\bigg(\frac{-s_{34}}{\mu^2}\bigg)
        + \log\bigg(\frac{-s_{23}}{\mu^2}\bigg) - \log\bigg(\frac{-s_{12}}{\mu^2}\bigg)
        + \log\bigg(\frac{-s_{14}}{\mu^2}\bigg)\!\bigg)\,.
    \label{eq:poles:catani:nf}
\end{align}
This subtraction scheme agrees with what was used in previous studies of 
non-factorisable contributions to VBF~\cite{Asteriadis:2023nyl} and is equivalent to the multiplicative minimal subtraction using a 
$Z$-factor~\cite{Becher:2009cu}.
In our explicit calculation we find that indeed $A^{(1)}_{\lambda,\textrm{fin}} = \mathcal{O}(\epsilon^0)$, which is an important consistency check of our calculation.

In detail, we have
\begin{align}
\label{eq:basisfunctions:explicit}
\begin{alignedat}[b]{9}
    f_{1,2}&\equiv-I_{12,\textrm{x}243}^{(1)}\,,&&\qquad & f_{1,3}&\equiv
        -I_{12,\textrm{id}}^{(1)}\,,&&\qquad & f_{1,4}&\equiv
        -I_{12,\textrm{x}345}^{(1)}\,,\\
    f_{1,9}&\equiv-I_{12,\textrm{x}45}^{(1)}\,,&&\qquad & f_{1,10}&\equiv
        -I_{12,\textrm{x}23}^{(1)}\,,&&\qquad & f_{1,11}&\equiv
        -I_{12,\textrm{x}23\textrm{x}45}^{(1)} \, ,
\end{alignedat}
\end{align}
where $I_{k,\sigma}^{(w)}$ is the $\epsilon^w$-coefficient of the master integral $I_k$ in permutation $\sigma$, which we write in the cycle notation of \texttt{Reduze 2}.
We note again that here we have already identified $\mu_0$ with $\mu$.

We write the finite remainder of the one-loop amplitude as
\begin{align}
\label{eq:ampl:1loop:final}
    A_{\lambda,\textrm{fin}}^{(1),\textrm{f}/\textrm{nf}} 
        = \Phi_\lambda \sum_{i,j} \epsilon^{j} \, 
        c^{(j),\textrm{f}/\textrm{nf}}_{\lambda,i} \, 
        G^{(j),\textrm{f}/\textrm{nf}}_{\lambda,i} \, ,
\end{align}
where we factored out the little-group weight $\Phi_\lambda$ as given in 
eq.~\eqref{eq:proj:spinorfactors}. The remaining parts are a little-group invariant series in 
$\epsilon$, where the coefficients $c^{(j)}_{\lambda,i}$ are rational functions 
in kinematic invariants, and  $G^{(j)}_{\lambda,i}$ are polynomials in 
transcendental basis functions, sometimes together with factors of (inverse) 
square roots and $\mathrm{tr}_5$. Note that the functions proportional to 
$\mathrm{tr}_5$ are parity-odd, while the others are parity-even.

For the projector method, the factor $\Phi_\lambda$, that carries the little-group weight, is divided out from the start and we make an ansatz in terms of invariants for the remainder, cf.\ eqs.~(\ref{eq:ampl:ff}~-~\ref{eq:ampl:ff4}).
By construction, the result of this computation is in the form of 
eq.~\eqref{eq:ampl:1loop:final}, and we only have to identify the factors in the sum.

Using the spinor-helicity method, we obtain a result in terms of invariants and spinor products that has the same transformation properties as $\Phi_\lambda$, but is not obviously proportional to it.
We divide the expression by $\Phi_\lambda$ and express the remaining spinor products in terms of invariants.
We first collect them term-by-term in the numerators via insertion of unities, e.g. $1 = \langle ij \rangle / \langle ij \rangle$, such that we can use $\langle ij \rangle [ij] = -2\, 
p_i\cdot p_j$ in the denominators. Since we already divided out $\Phi_\lambda$, that contains the 
little-group weight, the numerators have to be polynomials in six helicity 
products,
\begin{align}
\label{eq:spinorprods}
\begin{aligned}[b]
    \langle 12 \rangle & \langle 34 \rangle \, ,  & \langle 13 \rangle & \langle 
        24 \rangle \, ,  &\langle 14 \rangle &\langle 23 \rangle \, , \\
    [12] & [34]  \, ,  & [13] & [24] \, , & [14] & [23] \, ,
\end{aligned}
\end{align}
kinematic invariants, and $\textrm{tr}_5$. Using the Schouten 
identity~\eqref{eq:schouten}, we can replace one product in each row of eq.~\eqref{eq:spinorprods} with a 
linear combination of the remaining products in that row, and we chose to 
eliminate the last column. Two further ones can be rewritten using the 
pseudo-scalar five-point invariant~\eqref{eq:pseudoscalar},
\begin{align}
    \langle 13 \rangle \langle 24 \rangle  &= \bigg( \frac{s_{13} \, 
        s_{24} - s_{14} \, s_{23} + s_{12} \, s_{34}
        -\mathrm{tr}_5}{2 \, s_{12} \, s_{34}} \bigg) 
        \langle 12 \rangle \langle 34 \rangle\, , \nonumber\\
    [13][24]  &= \bigg( \frac{s_{13} \, s_{24} - s_{14} \, s_{23} 
        + s_{12} \, s_{34}+\mathrm{tr}_5}{2\,s_{12} \, s_{34}} \bigg) 
        [12][34]  \, .
\end{align}
After this step, spinor products can only appear in the numerators in the 
combination $\langle 12\rangle\langle 34\rangle[12][34] = s_{12}s_{34}$. 
The result from the spinor-helicity method is now in a similar form as the result from the projector method: master integrals with rational functions as coefficients. In both cases we continue in the same way.

Technically, our results are already of the form of eq.~\eqref{eq:ampl:1loop:final}, but further simplifications are possible.
We work order-by-order in $\epsilon$ and simplify rational coefficients of master integrals with a multivariate partial fraction 
decomposition, using the package \texttt{MultivariateApart} \cite{Heller:2021qkz}.
We then detect linear relations between the partial-fractioned expressions and promote a linearly independent set thereof to the coefficients $c^{(j)}_{\lambda,i}$. These then multiply linear combinations of basis functions $G^{(j)}_{\lambda,i}$, arriving at eq.~\eqref{eq:ampl:1loop:final}.


\section{Results and checks}
\label{sec:numeric}

We begin the discussion of our results with a comparison of the Born amplitude in all considered schemes. Both the projector computation as well as a direct calculation with BMHV-$\gamma_5$ give
the same result, which agrees with its four-dimensional limit
\begin{align}
    \mathcal{A}^{(0),\textrm{4D}}_{--} &= \mathcal{A}^{(0),\textrm{pr}}_{--} 
        = \mathcal{A}^{(0),\textrm{BMHV}}_{--} = +2 e^2 \, g_\textrm{vvh} \, 
        g_{q_1}^- \, g_{q_2}^- \frac{\langle 12\rangle[34]}{(s_{13}
        -m_\textrm{v}^2)(s_{24}-m_\textrm{v}^2)} \, , \\
    \mathcal{A}^{(0),\textrm{4D}}_{-+} &= \mathcal{A}^{(0),\textrm{pr}}_{-+} 
        = \mathcal{A}^{(0),\textrm{BMHV}}_{-+} = -2 e^2 \, g_\textrm{vvh} \, 
        g_{q_1}^- \, g_{q_2}^+ \frac{\langle 14\rangle[23]}{(s_{13}
        -m_\textrm{v}^2)(s_{24}-m_\textrm{v}^2)} \, , \\
    \mathcal{A}^{(0),\textrm{4D}}_{+-} &= \mathcal{A}^{(0),\textrm{pr}}_{+-} 
        = \mathcal{A}^{(0),\textrm{BMHV}}_{+-} = -2 e^2 \, g_\textrm{vvh} \, 
        g_{q_1}^+ \, g_{q_2}^- \frac{\langle 23\rangle[14]}{(s_{13}
        -m_\textrm{v}^2)(s_{24}-m_\textrm{v}^2)} \, , \\
    \mathcal{A}^{(0),\textrm{4D}}_{++} &= \mathcal{A}^{(0),\textrm{pr}}_{++} 
        = \mathcal{A}^{(0),\textrm{BMHV}}_{++} = +2 e^2 \, g_\textrm{vvh} \, 
        g_{q_1}^+ \, g_{q_2}^+ \frac{\langle 34\rangle[12]}{(s_{13}
        -m_\textrm{v}^2)(s_{24}-m_\textrm{v}^2)} \, .
\end{align}
In a direct computation with anticommuting $\gamma_5$, we obtain an additional 
evanescent term. We find
\begin{align}
    \label{eq:res:born:ac:mm}
    \frac{\mathcal{A}^{(0),\textrm{ac}}_{--}}
         {\mathcal{A}^{(0),\textrm{4D}}_{--}} &= 
        1 - \epsilon \frac{[13][24]}{[12][34]} = 1 - \epsilon\frac{s_{13} 
        \, s_{24} - s_{14} \, s_{23} + s_{12} \, s_{34}+\mathrm{tr}_5}{2\,s_{12} \, 
        s_{34}} \, , \\
    \frac{\mathcal{A}^{(0),\textrm{ac}}_{-+}}
         {\mathcal{A}^{(0),\textrm{4D}}_{-+}} &= 
        1 - \epsilon \frac{\langle 13 \rangle \langle 24 \rangle}{\langle 14 
        \rangle \langle 23 \rangle} = 1 - \epsilon\frac{s_{13} \, s_{24} 
        - s_{12} \, s_{34} + s_{14} \, s_{23}+\mathrm{tr}_5}{2\,s_{14} \, 
        s_{23}} \, , \\
    \frac{\mathcal{A}^{(0),\textrm{ac}}_{+-}}
         {\mathcal{A}^{(0),\textrm{4D}}_{+-}} &= 
        1 - \epsilon \frac{[13][24]}{[14][23]}  = 1 - \epsilon\frac{s_{13} \, 
        s_{24} - s_{12} \, s_{34} + s_{14} \, s_{23}-\mathrm{tr}_5}{2\,s_{14} \, 
        s_{23}} \, , \\
    \label{eq:res:born:ac:pp}
    \frac{\mathcal{A}^{(0),\textrm{ac}}_{++}}
         {\mathcal{A}^{(0),\textrm{4D}}_{++}} &= 
        1 - \epsilon \frac{\langle 13 \rangle \langle 24 \rangle}{\langle 12 
        \rangle \langle 34 \rangle}  = 1 - \epsilon\frac{s_{13} \, s_{24} 
        - s_{14} \, s_{23} + s_{12} \, s_{34}-\mathrm{tr}_5}{2\,s_{12} \, 
        s_{34}} \, .
\end{align}
We recall that the $\mathcal{O}(\epsilon)$-term in the anticommuting $\gamma_5$ 
scheme depends on the choice of the momentum $k$, where we chose $k=p_1$. In 
sections~\ref{sec:helamp:anticom}-\ref{sec:helamp:thooft-veltman}, this discrepancy 
between the spinor-helicity computations using an anticommuting and a 
BMHV $\gamma_5$ has been traced back to the fact that 
eq.~\eqref{eq:completeness:anti} is a prescription, while 
eq.~\eqref{eq:completeness:BMHV} is an identity. The projector computation does 
not lead to $\mathcal{O}(\epsilon)$-terms because of our choice of tensor 
structures~\eqref{eq:proj:tensors}.

\begin{figure}[t]
    \centering
    \vspace{15pt}
    \includegraphics[height=115pt]{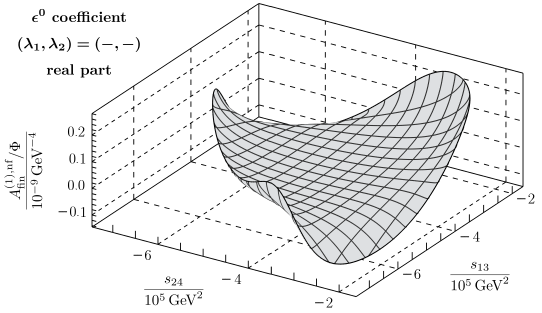}
    \hspace{10pt}
    \includegraphics[height=115pt]{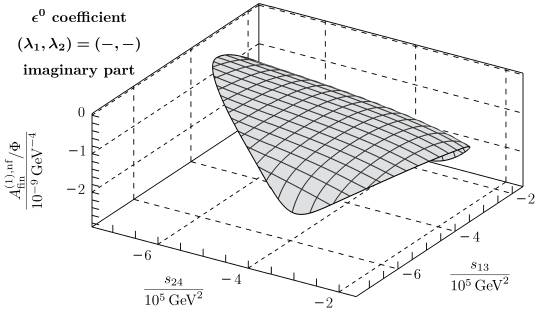} \\[5pt]
    \includegraphics[height=115pt]{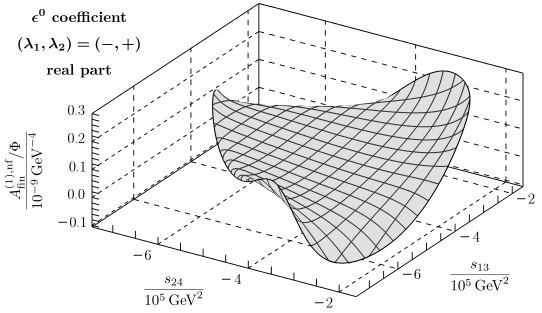}
    \hspace{10pt}
    \includegraphics[height=115pt]{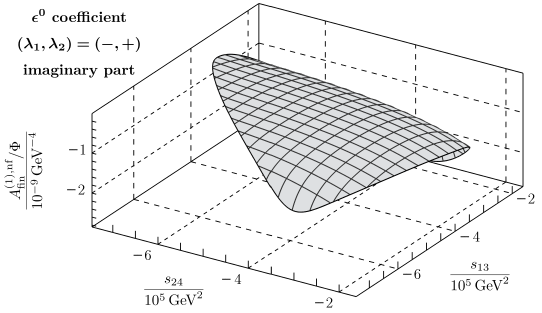} 
    \caption{Real (left) and imaginary (right) part of 
    $\mathcal{O}(\epsilon^0)$-coefficients of $A_{\lambda_1\lambda_2, 
    \textrm{fin}}^{(1),\textrm{nf}}/\Phi_{\lambda_1\lambda_2}$ for helicities 
    $(\lambda_1,\lambda_2) = (-,-)$ (upper) and $(\lambda_1,\lambda_2) = (-,+)$ 
    (lower) as a function of $s_{13}$ and $s_{24}$. Other variables are kept at 
    fixed values, see the text for details.}
    \label{fig:results:finite}
\end{figure}

\begin{figure}[t]
    \centering
    \vspace{15pt}
    \includegraphics[height=115pt]{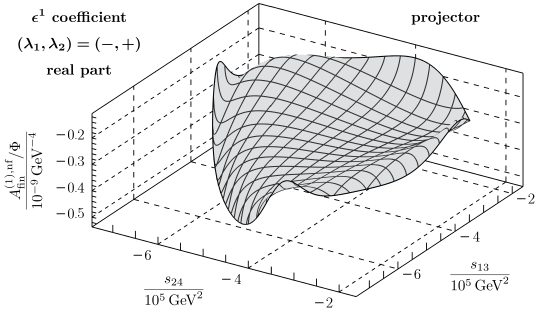}
    \hspace{10pt}
    \includegraphics[height=115pt]{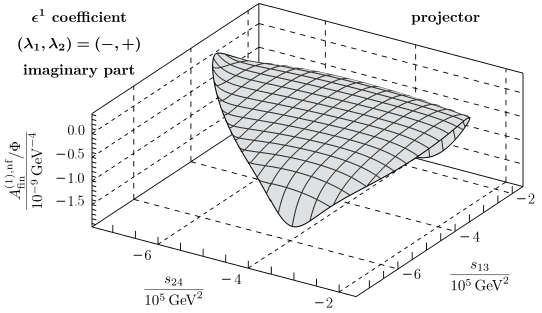} \\[5pt]
    \includegraphics[height=115pt]{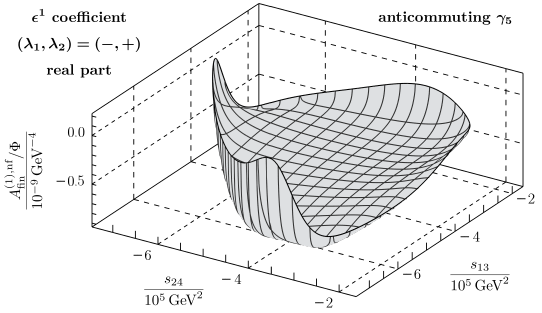}
    \hspace{10pt}
    \includegraphics[height=115pt]{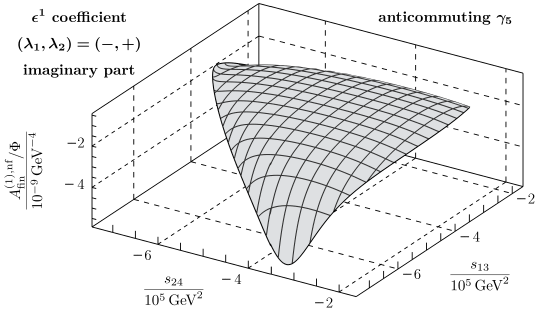} 
    \caption{Comparison of $\mathcal{O}(\epsilon)$ coefficients of 
    $A_{-+,\textrm{fin}}^{(1),\textrm{nf}}/\Phi_{-+}$ computed using the 
    projector method (top), and spinor-helicity methods with an anticommuting 
    $\gamma_5$ scheme (bottom). We show real parts (left) and imaginary parts 
    (right) separately.}
    \label{fig:results:ep1}
\end{figure}

Next, we discuss our one-loop results.
The fact that finite remainders are indeed regular for $\epsilon\to 0$ after subtraction of the predicted infrared poles~(\ref{eq:poles:catani:f}~-~\ref{eq:poles:catani:nf}) provides a 
powerful check of our amplitudes.
Our results for the finite remainder for all three schemes considered in this paper agree for the $\epsilon^0$ term and start to differ for the $\epsilon^1$ term.

The complete one-loop results are too lengthy to be displayed here.
We present the factorisable and non-factorisable one-loop finite remainders up to 
$\mathcal{O}(\epsilon^2)$ in appendix~\ref{app:ampl:1loop} and ancillary files, 
respectively.
Here and in the following, we restrict ourselves to an evaluation in two different schemes: the projector method and the helicity method with anticommuting $\gamma_5$.

The master integrals have been implemented in a \texttt{C++} library, which was 
validated against \texttt{PySecDec}~\cite{Borowka:2017idc} and up to the finite 
part against \texttt{LoopTools}~\cite{Hahn:2000jm}. This allows us to evaluate the amplitude numerically beyond $\mathcal{O}(\epsilon^0)$.
To study 
it further, we consider a physical crossing,
\begin{align}
\label{eq:process:physical}
    \bar{q}_1(-p_1) + \bar{q}_2(-p_2) \rightarrow \bar{q}_1^\prime(p_3) + 
        \bar{q}_2^\prime(p_4) + h(p_5) \, .
\end{align}
In appendix~\ref{app:benchmark} we give numerical results at a benchmark 
phase-space point. The VBF one-loop amplitudes were computed to 
$\mathcal{O}(\epsilon^0)$ in ref.~\cite{Campanario:2013fsa}. These results where 
adapted to provide non-factorisable amplitudes for 
ref.~\cite{Asteriadis:2023nyl}, and we use this implementation to validate our 
$\mathcal{O}(\epsilon^0)$-contributions.
We find agreement for the helicity-summed and colour-stripped amplitude squared to about 15 digits for generic points in phase space.

To further illustrate our results, we consider two-dimensional slices of the physical phase space. For 
centre-of-mass energies $s_{12}\equiv s > m_\textrm{h}^2$, the physical region 
for the crossing~\eqref{eq:process:physical} is given by
\begin{gather}
\label{eq:PhysicalRegion}
    s_{34} > 0 \, , \ s_{24} < 0 \, , \ s_{14} < 0 \, , \ s_{13} < 0 \, , 
        s_{14} + s_{24} + s_{34} < 0 \, , \ s_{13} + s_{14} + s_{34} < 0 \, , \nonumber\\ 
    - (s_{13} + s_{14}) < s - m_\textrm{h}^2 \, , \ - (s_{14} + s_{24}) < 
        s - m_\textrm{h}^2  \, , \nonumber\\
    -(s_{13} + s_{14} + s_{24} + s_{34}) < s - m_\textrm{h}^2\, , \nonumber\\
    \Delta_5 < 0 \, ,
\end{gather}
with the Gram determinant~\eqref{eq:gramdet}. For our plots, we choose $\sqrt{s} 
= 1 \, \textrm{TeV}$, $m_\textrm{h} = 125 \, \textrm{GeV}$, a typical mass 
$\sqrt{s_{34}} = 650 \, \textrm{GeV}$ of the dijet system in VBF, and 
$\sqrt{-s_{14}} = 500 \, \textrm{GeV}$. Moreover, $m_\textrm{v}^2 = 
m_\textrm{Z}^2 - \mathrm{i} \, m_\textrm{Z} \,\Gamma_\textrm{Z}$ with 
$m_\textrm{Z} = 91.1876 \, \mathrm{GeV}$, $\Gamma_\textrm{Z} = 2.4952 \, 
\mathrm{GeV}$ and $\mu^2 = m_\textrm{Z}^2$. For $\mathrm{Im}(\mathrm{tr}_5)>0$, 
real and imaginary parts for different helicity combinations are shown in 
figure~\ref{fig:results:finite} as functions of the remaining variables $s_{13}$ 
and $s_{24}$.

At higher orders in $\epsilon$, the finite remainders are scheme-dependent.
We compare the real and imaginary parts of the $\epsilon^1$-coefficient of 
$A_{-+,\textrm{fin}}^{(1),\textrm{nf}}/\Phi_{-+}$ for both calculational schemes in 
figure~\ref{fig:results:ep1}.
In the plots, numerical differences between the two schemes are clearly visible.


\section{Conclusion}
\label{sec:conclusion}

In this work, we presented an analytical calculation of helicity amplitudes for 
vector-boson fusion at one-loop order in QCD. We included both factorising and 
non-factorisable corrections. In contrast to previous 
calculations~\cite{Figy:2003nv, Figy:2004pt, Berger:2004pca}, we included 
contributions up to $\mathcal{O}(\epsilon^2)$. The $\mathcal{O}(\epsilon)$ terms 
constitute necessary input for the finite remainder of the non-factorisable 
two-loop contributions. We also provide the $\mathcal{O}(\epsilon^2)$ 
coefficients to enable applications in more general contexts, including 
electromagnetic corrections, where this term is required. Our analytical results 
are expressed in terms of an algebraically independent set of transcendental 
basis functions and linearly independent rational coefficients, allowing for a 
compact representation of the amplitudes.

To compute the amplitude, we adapted the spinor-helicity formalism to 
dimensional regularisation and investigated the interplay of such a procedure 
with the choice of the $\gamma_5$ scheme. We obtained the complete one-loop 
finite remainder using  the spinor-helicity method both with anticommuting 
$\gamma_5$ and with the Breitenlohner-Maison~/~'t Hooft-Veltman scheme 
for $\gamma_5$, as well as using a projector method. We find agreement between 
these schemes at $\mathcal{O}(\epsilon^0)$, but differences at 
$\mathcal{O}(\epsilon^1)$ and higher.


\acknowledgments

We gratefully acknowledge useful discussions with Ajjath Abdul Hameed, Bakul Agarwal and Stephen Jones.
We are thankful to Terrance Figy for the repeated use of his Fortran code to compute the 
non-factorisable one-loop amplitudes.

\appendix


\section{Prefactors in canonical basis}
\label{app:basisdefpolys}

Here, we define the polynomials $Q_i$ in eq.~\eqref{eq:MI:canonical-basis}. They 
are given by
\begin{align}
    Q_0 &= -2\, \bar{s}_{34} \big(m^4 \bar{s}_{12} \bar{s}_{15} - m^4 \bar{s}_{12} \bar{s}_{23} + m^4 \bar{s}_{15} 
        \bar{s}_{23} - m^2  \bar{s}_{12} \bar{s}_{15} \bar{s}_{23} - m^4 \bar{s}_{23}^2 \nonumber\\
        & \hspace{11pt} + m^2  \bar{s}_{12} \bar{s}_{23}^2 - m^4 \bar{s}_1  \bar{s}_{34} + m^2  \bar{s}_1  
        \bar{s}_{23} \bar{s}_{34} - m^2  \bar{s}_{23}^2 \bar{s}_{34} + m^4 \bar{s}_{12} \bar{s}_{45} - m^4 \bar{s}_{15} 
        \bar{s}_{45} \nonumber\\
        & \hspace{11pt} - m^2  \bar{s}_{12} \bar{s}_{15} \bar{s}_{45}  + 2\, m^4 \bar{s}_{23} \bar{s}_{45}  
        - m^2  \bar{s}_{12} \bar{s}_{23} \bar{s}_{45}  - m^2  \bar{s}_{15} \bar{s}_{23} \bar{s}_{45}+ \bar{s}_{12} \bar{s}_{15}
        \bar{s}_{23} \bar{s}_{45}  \nonumber\\
        & \hspace{11pt}+ m^2  \bar{s}_1  \bar{s}_{34} \bar{s}_{45} + 2\, m^2  \bar{s}_{23} \bar{s}_{34} \bar{s}_{45}
        - \bar{s}_1  \bar{s}_{23} \bar{s}_{34} \bar{s}_{45} - m^4 \bar{s}_{45}^2 + m^2  \bar{s}_{15} \bar{s}_{45}^2\nonumber\\
        & \hspace{11pt} - m^2  \bar{s}_{34} \bar{s}_{45}^2\big) \,,
    \\
    Q_1 &= \bar{s}_{34} \big(-m^2 \bar{s}_{12} \bar{s}_{15} + m^2 \bar{s}_{12} \bar{s}_{23} - 2\, m^2 \bar{s}_{15} 
        \bar{s}_{23} + \bar{s}_{12} \bar{s}_{15} \bar{s}_{23} + 2\, m^2  \bar{s}_{23}^2 - \bar{s}_{12} \bar{s}_{23}^2  \nonumber\\
        & \hspace{11pt}  + m^2  \bar{s}_1  \bar{s}_{34} + m^2  \bar{s}_{23} \bar{s}_{34} - \bar{s}_1  \bar{s}_{23} 
        \bar{s}_{34} + \bar{s}_{23}^2 \bar{s}_{34} + m^2 \bar{s}_{15} \bar{s}_{45} - 2\, m^2 \bar{s}_{23} \bar{s}_{45} \nonumber\\
        & \hspace{11pt}  + \bar{s}_{15} \bar{s}_{23} \bar{s}_{45} - m^2  \bar{s}_{34} \bar{s}_{45} - \bar{s}_{23}
        \bar{s}_{34} \bar{s}_{45}\big) \,,
    \\
    Q_2 &= \bar{s}_{34} \big(-m^2 \bar{s}_{12} \bar{s}_{15} + m^2 \bar{s}_{12} \bar{s}_{23} + m^2 \bar{s}_1 \bar{s}_{34} 
        - m^2  \bar{s}_{23} \bar{s}_{34} - 2\, m^2  \bar{s}_{12} \bar{s}_{45} + m^2  \bar{s}_{15} \bar{s}_{45}  \nonumber\\
        & \hspace{11pt} + \bar{s}_{12} \bar{s}_{15} \bar{s}_{45} - 2\, m^2  \bar{s}_{23} \bar{s}_{45} + \bar{s}_{12}
        \bar{s}_{23} \bar{s}_{45} + m^2  \bar{s}_{34} \bar{s}_{45} - \bar{s}_1  \bar{s}_{34} \bar{s}_{45} - \bar{s}_{23} \bar{s}_{34} 
        \bar{s}_{45}\nonumber\\
        & \hspace{11pt}  + 2\, m^2  \bar{s}_{45}^2 - \bar{s}_{15} \bar{s}_{45}^2 + \bar{s}_{34} 
        \bar{s}_{45}^2\big) \,, 
    \\
    Q_3 &= m^2  \bar{s}_{12} \bar{s}_{15}^2 - 2\, m^2  \bar{s}_{12} \bar{s}_{15} \bar{s}_{23} + m^2  \bar{s}_{12}
        \bar{s}_{23}^2 - m^2  \bar{s}_1  \bar{s}_{15} \bar{s}_{34} + m^2  \bar{s}_1  \bar{s}_{23} \bar{s}_{34} \nonumber\\
        & \hspace{11pt} + m^2  \bar{s}_{15} \bar{s}_{23} \bar{s}_{34} - m^2  \bar{s}_{23}^2 \bar{s}_{34} 
        + m^2  \bar{s}_{12} \bar{s}_{15} \bar{s}_{45} - m^2  \bar{s}_{15}^2 \bar{s}_{45} - \bar{s}_{12} 
        \bar{s}_{15}^2 \bar{s}_{45} \nonumber\\
        & \hspace{11pt}  - m^2  \bar{s}_{12} \bar{s}_{23} \bar{s}_{45} + m^2  \bar{s}_{15} \bar{s}_{23} 
        \bar{s}_{45} + \bar{s}_{12} \bar{s}_{15} \bar{s}_{23} \bar{s}_{45}  + m^2  \bar{s}_1  \bar{s}_{34} \bar{s}_{45} 
        - m^2  \bar{s}_{15} \bar{s}_{34} \bar{s}_{45} \nonumber\\
        & \hspace{11pt}  + \bar{s}_1  \bar{s}_{15} \bar{s}_{34} \bar{s}_{45} + 2\, m^2  \bar{s}_{23} \bar{s}_{34}
        \bar{s}_{45} - 2\, \bar{s}_1  \bar{s}_{23} \bar{s}_{34} \bar{s}_{45} + \bar{s}_{15} \bar{s}_{23} \bar{s}_{34} \bar{s}_{45}
        - m^2  \bar{s}_{15} \bar{s}_{45}^2  \nonumber\\
        & \hspace{11pt}  + \bar{s}_{15}^2 \bar{s}_{45}^2 - m^2  \bar{s}_{34} \bar{s}_{45}^2 - \bar{s}_{15}
        \bar{s}_{34} \bar{s}_{45}^2\,, 
    \\
    Q_4 &= -m^2  \bar{s}_{12}^2 \bar{s}_{15} - m^2  \bar{s}_{12} \bar{s}_{15}^2 + \bar{s}_{12}^2 \bar{s}_{15}^2 
        + m^2  \bar{s}_{12}^2 \bar{s}_{23} +    m^2  \bar{s}_{12} \bar{s}_{15} \bar{s}_{23} - \bar{s}_{12}^2 
        \bar{s}_{15} \bar{s}_{23} \nonumber\\
        & \hspace{11pt} + m^2  \bar{s}_1  \bar{s}_{12} \bar{s}_{34} + m^2  \bar{s}_1  \bar{s}_{15} \bar{s}_{34} 
        + 2\, m^2  \bar{s}_{12} \bar{s}_{15} \bar{s}_{34} - 2\, \bar{s}_1  \bar{s}_{12} \bar{s}_{15} \bar{s}_{34} 
        - 2\, m^2  \bar{s}_1  \bar{s}_{23} \bar{s}_{34} \nonumber\\
        & \hspace{11pt} - m^2  \bar{s}_{12} \bar{s}_{23} \bar{s}_{34} + \bar{s}_1  \bar{s}_{12} \bar{s}_{23} 
        \bar{s}_{34} + m^2  \bar{s}_{15} \bar{s}_{23} \bar{s}_{34} + \bar{s}_{12} \bar{s}_{15} \bar{s}_{23} \bar{s}_{34} 
        - 2\, m^2  \bar{s}_1  \bar{s}_{34}^2 \nonumber\\
        & \hspace{11pt} + \bar{s}_1 ^2 \bar{s}_{34}^2- \bar{s}_1  \bar{s}_{23} \bar{s}_{34}^2 + m^2 
        \bar{s}_{12} \bar{s}_{15} \bar{s}_{45} + m^2  \bar{s}_{15}^2 \bar{s}_{45} -  \bar{s}_{12} \bar{s}_{15}^2
        \bar{s}_{45} - 2\, m^2  \bar{s}_1  \bar{s}_{34} \bar{s}_{45} \nonumber\\
        & \hspace{11pt} + m^2  \bar{s}_{12} \bar{s}_{34} \bar{s}_{45} - m^2  \bar{s}_{15} \bar{s}_{34} 
        \bar{s}_{45} + \bar{s}_1  \bar{s}_{15} \bar{s}_{34} \bar{s}_{45} + \bar{s}_{12} \bar{s}_{15} \bar{s}_{34} \bar{s}_{45} 
        - \bar{s}_1  \bar{s}_{34}^2 \bar{s}_{45}\,,
    \\
    Q_5 &= m^2  \bar{s}_{12}^2 \bar{s}_{15} - m^2  \bar{s}_{12}^2 \bar{s}_{23} + m^2  \bar{s}_{12} 
        \bar{s}_{15} \bar{s}_{23} - \bar{s}_{12}^2 \bar{s}_{15} \bar{s}_{23} - m^2  \bar{s}_{12} \bar{s}_{23}^2 
        + \bar{s}_{12}^2 \bar{s}_{23}^2  \nonumber\\
        & \hspace{11pt}- m^2  \bar{s}_1  \bar{s}_{12} \bar{s}_{34} + m^2  \bar{s}_1  \bar{s}_{23} \bar{s}_{34} 
        - m^2  \bar{s}_{12} \bar{s}_{23} \bar{s}_{34} + \bar{s}_1  \bar{s}_{12} \bar{s}_{23} \bar{s}_{34} - m^2  
        \bar{s}_{23}^2 \bar{s}_{34} \nonumber\\
        & \hspace{11pt} - \bar{s}_{12} \bar{s}_{23}^2 \bar{s}_{34} - 2\, m^2  \bar{s}_{12} \bar{s}_{15} 
        \bar{s}_{45}+ m^2  \bar{s}_{12} \bar{s}_{23} \bar{s}_{45}  -  m^2  \bar{s}_{15} \bar{s}_{23} \bar{s}_{45} 
        + \bar{s}_{12} \bar{s}_{15} \bar{s}_{23} \bar{s}_{45} \nonumber\\
        & \hspace{11pt} + m^2  \bar{s}_1  \bar{s}_{34} \bar{s}_{45} + m^2  \bar{s}_{12} \bar{s}_{34}
        \bar{s}_{45} + 2\, m^2  \bar{s}_{23} \bar{s}_{34} \bar{s}_{45} - 2 \bar{s}_1  \bar{s}_{23} \bar{s}_{34} 
        \bar{s}_{45}  + \bar{s}_{12} \bar{s}_{23} \bar{s}_{34} \bar{s}_{45} \nonumber\\
        & \hspace{11pt}  + m^2  \bar{s}_{15} \bar{s}_{45}^2 - m^2  \bar{s}_{34} \bar{s}_{45}^2\,.
\end{align}


\section{Factorisable one-loop finite remainders}
\label{app:ampl:1loop}

The results for the finite remainders of the factorisable one-loop contributions 
can be expressed entirely in terms of the triangle integral $I_{12}$ with one 
off-shell leg and massless propagators, and thus in terms of the basis functions 
$f_{1,9}$ and $f_{1,10}$, see eq.~\eqref{eq:basisfunctions:explicit}. Here, we 
present our results for the quantities $c_{\lambda,i}^{(j)}$ and 
$G_{\lambda,i}^{(j)}$, from which the finite remainder can be built according to 
eq.~\eqref{eq:ampl:1loop:final}. We provide these results for two calculational 
schemes.

\paragraph{Spinor-helicity method with anticommuting \texorpdfstring{\boldmath $\gamma_5$}{gamma 5}.}

Using the spinor-helicity method with anticommuting $\gamma_5$ we find the 
rational coefficients
\begin{align}
\begin{aligned}[b]
    c_{\lambda,1}^{(0),\textrm{f},\textrm{ac}} &= 
        -\frac{1}{2 d_1 d_2} \, , & \\
    c_{\lambda,1}^{(1),\textrm{f},\textrm{ac}} &= 
        \frac{-3 + 2 c_\lambda}{4 d_1 d_2} \, ,  &
    c_{\lambda,2}^{(1),\textrm{f},\textrm{ac}} &= 
        \frac{s_\lambda}{4 s_{14}  s_{23}d_1 d_2} \, ,  &
    c_{\lambda,3}^{(1),\textrm{f},\textrm{ac}} &= 
        \frac{1}{6d_1 d_2} \, , & \\
    c_{\lambda,1}^{(2),\textrm{f},\textrm{ac}} &= 
        -\frac{1}{3} c_{\lambda,1}^{(1),\textrm{f},\textrm{ac}} \, ,  &
    c_{\lambda,2}^{(2),\textrm{f},\textrm{ac}} &= 
        -\frac{1}{3} c_{\lambda,2}^{(1),\textrm{f},\textrm{ac}} \, ,  &
    c_{\lambda,3}^{(2),\textrm{f},\textrm{ac}} &= 
        -\frac{1}{4} c_{\lambda,3}^{(1),\textrm{f},\textrm{ac}} \, , &
\end{aligned}
\end{align}
where $d_1 \equiv (s_{13}-m_\textrm{v}^2)$ and $d_2 \equiv 
(s_{24}-m_\textrm{v}^2)$ are inverse propagators and
\begin{align}
    c_{--} &= c_{++} = \frac{s_{13} \, s_{24} - s_{14} \, s_{23} 
        + s_{12} \, s_{34}}{2\,s_{12} \, s_{34}} \, , &
    s_{--} &= s_{-+} = 1 \, , \nonumber\\    
    c_{-+} &= c_{+-} = \frac{s_{13} \, s_{24} - s_{12} \, s_{34} 
        + s_{14} \, s_{23}}{2\,s_{14} \, s_{23}} \, , &
    s_{+-} &= s_{++} = -1 \, .
\end{align}
We note that $c_\lambda$ are the evanescent terms, apart from the 
$\textrm{tr}_5$ term that we choose to include in the transcendental functions, 
cf.\ eqs.~(\ref{eq:res:born:ac:mm})~-~(\ref{eq:res:born:ac:pp}). Signs of the 
$\textrm{tr}_5$ terms are absorbed into $s_\lambda$. These contributions appear 
because we factor out the little group weight $\Phi_\lambda$ of the 
four-dimensional Born amplitude and omit the $\mathcal{O}(\epsilon)$ evanescent 
contributions.
We note that not omitting them leads to a much simpler result.

The transcendental functions are given by
\begin{align}
    G_{\lambda,1}^{(0),\textrm{f},\textrm{ac}} &= 16 - 2 \zeta_2 - 3 f_{1,9} 
        + f_{1,9}^2 - 3 f_{1,10} + f_{1,10}^2 \, ,  \\[10pt]
    G_{\lambda,1}^{(1),\textrm{f},\textrm{ac}} &= 16 - 2 \zeta_2 - 3 f_{1,9} 
        + f_{1,9}^2 - 3 f_{1,10} + f_{1,10}^2  \, , \nonumber\\ 
    G_{\lambda,2}^{(1),\textrm{f},\textrm{ac}} &= 
        G_{\lambda,1}^{(1),\textrm{f},\textrm{ac}} \, \textrm{tr}_5 \, , \nonumber\\
    G_{\lambda,3}^{(1),\textrm{f},\textrm{ac}} &= -24 + 28 \zeta_3 
        + \frac{21}{2} f_{1,9} - 3 \zeta_2 f_{1,9} + f_{1,9}^3 + \frac{21}{2} 
        f_{1,10} - 3 \zeta_2 f_{1,10} + f_{1,10}^3 \, , 
    \\[10pt]
    G_{\lambda,1}^{(2),\textrm{f},\textrm{ac}} &= -96 -3 f_{1,9} \zeta_2
        -3 f_{1,10} \zeta_2+f_{1,9}^3 -\frac{9 }{2} f_{1,9}^2 +24 f_{1,9}
        +f_{1,10}^3+24 f_{1,10}\nonumber\\
        &\hspace{11pt} -\frac{9 }{2} f_{1,10}^2+9 \zeta_2+28 \zeta_3 \, , \nonumber\\
    G_{\lambda,2}^{(2),\textrm{f},\textrm{ac}} &= 
        G_{\lambda,1}^{(2),\textrm{f},\textrm{ac}} \, \textrm{tr}_5 \, , \nonumber\\
    G_{\lambda,3}^{(2),\textrm{f},\textrm{ac}} &= 192 -6 f_{1,9}^2 \zeta_2
        -6 f_{1,10}^2 \zeta_2+56 f_{1,9} \zeta_3+56 f_{1,10} \zeta_3
        +f_{1,9}^4+21 f_{1,9}^2 \nonumber\\
        &\hspace{11pt} -48 f_{1,9}+f_{1,10}^4+21 f_{1,10}^2-48 f_{1,10}
        -\frac{282}{5} \zeta_2^2 -42 \zeta_2 \, .
\end{align}

\paragraph{Projector method.}

With the projector method, we obtain
\begin{align}
    c_{\lambda,1}^{(j),\textrm{f},\textrm{pr}} &= 
        \frac{1}{d_1 d_2}\,,\\
    G_{\lambda,1}^{(0),\textrm{f},\textrm{pr}} &= -8+\zeta_2
        +\frac{3}{2}\,f_{1,9}-\frac{1}{2}\,f_{1,9}^2+\frac{3}{2}\,f_{1,10}
        -\frac{1}{2}\,f_{1,10}^2,\\
    G_{\lambda,1}^{(1),\textrm{f},\textrm{pr}} &= -16+\frac{3}{2}\,\zeta_2
        +\frac{14}{3}\,\zeta_3+4\, f_{1,9}-\frac{1}{2} \zeta_2 f_{1,9}
        -\frac{3}{4}\,f_{1,9}^2+\frac{1}{6}f_{1,9}^3 \nonumber\\
        &\hspace{11pt}+4\, f_{1,10}-\frac{1}{2}\, \zeta_2 f_{1,10}
        - \frac{3}{4}\,f_{1,10}^2+\frac{1}{6}\,f_{1,10}^3\,,
    \\
    G_{\lambda,1}^{(2),\textrm{f},\textrm{pr}} &= -32+4\, \zeta_2
        +\frac{47}{20}\,\zeta_2^2 +7\, \zeta_3+8\, f_{1,9} - \frac{3}{4}\, 
        \zeta_2 f_{1,9}- \frac{7}{3}\, \zeta_3 f_{1,9} \nonumber\\
        & \hspace{11pt}- 2\, f_{1,9}^2 + \frac{1}{4}\, \zeta_2 f_{1,9}^2
        +\frac{1}{4}\,f_{1,9}^3 - \frac{1}{24}\,f_{1,9}^4 + 8\, f_{1,10} 
        - \frac{3}{4}\, \zeta_2 f_{1,10} \nonumber\\
        & \hspace{11pt}-\frac{7}{3}\, \zeta_3 f_{1,10} - 2\, f_{1,10}^2 
        + \frac{1}{4}\, \zeta_2 f_{1,10}^2 + \frac{1}{4}\,f_{1,10}^3 
        - \frac{1}{24}\,f_{1,10}^4\,.
\end{align}


\section{Numerical results}
\label{app:benchmark}

In this appendix, we give numerical values for the finite remainders at a physical phase-space point, cf.\ eq.~\eqref{eq:PhysicalRegion}. We use
\begin{align}
\begin{aligned}[b]
    -p_1 &= (1120.2069, 0, 0,  1120.2069) \, \mathrm{GeV} \, , \\
    -p_2 &= (376.73539, 0, 0, -376.73539) \, \mathrm{GeV} \, , \\
     p_3 &= (307.29587,  21.658129,  16.682320,  306.07740) \, 
        \mathrm{GeV} \, , \\
     p_4 &= (372.40019, -34.550737,  50.598215, -367.32543) \, 
        \mathrm{GeV} \, , \\
     p_5 &= (817.24619,  12.892608, -67.280535,  804.71949) \, 
        \mathrm{GeV} \, , \\
    m_\textrm{Z} &= 91.1876 \, \mathrm{GeV} \, , \\
    \Gamma_\textrm{Z} &= 2.4952 \, \mathrm{GeV} \, , \\
    m_\textrm{v}^2 &= m_\textrm{Z}^2 - \mathrm{i} \, m_\textrm{Z} \, 
        \Gamma_\textrm{Z} \, , \\
    \mu^2 &= m_\textrm{Z}^2 \, .
\end{aligned}
\label{eq:benchmark:point}
\end{align}
Results for Born amplitudes are given in table~\ref{tab:benchmark:born}, for 
factorisable one-loop amplitudes in table~\ref{tab:benchmark:fact} and for 
non-factorisable one-loop amplitudes in table~\ref{tab:benchmark:nfact}, where we 
factored out colour structures and coupling constants according to
eqs.~(\ref{eq:ampl:born:A}~-~\ref{eq:ampl}).

\begin{table}[!ht]
\begin{center}
\small
\begin{tabular}{cccc}
    \toprule
    $A_{\lambda}^{(0),\textrm{pr}}
        /\textrm{GeV}^{-2}$ \hspace{-25pt} & 
        $\epsilon^0$ & $\epsilon^1$ & $\epsilon^2$ \\
    \midrule
    $(-,-)$ & $-0.0069689 - 0.0110303 \, \mathrm{i}$ & $0$ & $0$ \\
    $(-,+)$ & $-0.0077773 + 0.0104794 \, \mathrm{i}$ & $0$ & $0$ \\
    $(+,-)$ & $-0.0069296 - 0.0110582 \, \mathrm{i}$ & $0$ & $0$ \\ 
    $(+,+)$ & $-0.0078142 + 0.0104485 \, \mathrm{i}$ & $0$ & $0$ \\
    \midrule
    $A_{\lambda}^{(0),\textrm{ac}}
        /\textrm{GeV}^{-2}$ \hspace{-25pt} & 
        $\epsilon^0$ & $\epsilon^1$ \\
    \midrule
    $(-,-)$ & $-0.0069689 - 0.0110303 \, \mathrm{i}$ &
              $+0.0000393 - 0.0000279 \, \mathrm{i}$ & $0$ \\
    $(-,+)$ & $-0.0077773 + 0.0104794 \, \mathrm{i}$ & 
              $-0.0000370 - 0.0000309 \, \mathrm{i}$ & $0$ \\
    $(+,-)$ & $-0.0069296 - 0.0110582 \, \mathrm{i}$ & 
              $-0.0000393 + 0.0000279 \, \mathrm{i}$ & $0$ \\
    $(+,+)$ & $-0.0078142 + 0.0104485 \, \mathrm{i}$ & 
              $+0.0000370 + 0.0000309 \, \mathrm{i}$ & $0$ \\
    \bottomrule
\end{tabular}
\end{center}
\caption{Born amplitudes at the point eq.~\eqref{eq:benchmark:point}, for the 
process eq.~\eqref{eq:process:physical}.}
\label{tab:benchmark:born}
\end{table}

\begin{table}[!ht]
\begin{center}
\small
\begin{tabular}{cccc}
    \toprule
    $A_{\lambda,\textrm{fin}}^{(1),\textrm{f},\textrm{pr}}
        /\textrm{GeV}^{-2}$ \hspace{-25pt} & 
        $\epsilon^0$ & $\epsilon^1$ & $\epsilon^2$ \\ 
    \midrule
    $(-,-)$ & $+0.0704773 + 0.1115518 \, \mathrm{i}$ & 
              $+0.1088695 + 0.1723194 \, \mathrm{i}$ & 
              $+0.1505906 + 0.2383558 \, \mathrm{i}$ \\
    $(-,+)$ & $+0.0786529 - 0.1059797 \, \mathrm{i}$ & 
              $+0.1214989 - 0.1637118 \, \mathrm{i}$ & 
              $+0.1680598 - 0.2264496 \, \mathrm{i}$ \\
    $(+,-)$ & $+0.0700799 + 0.1118337 \, \mathrm{i}$ & 
              $+0.1082557 + 0.1727548 \, \mathrm{i}$ & 
              $+0.1497415 + 0.2389581 \, \mathrm{i}$ \\
    $(+,+)$ & $+0.0790269 - 0.1056674 \, \mathrm{i}$ & 
              $+0.1220766 - 0.1632294 \, \mathrm{i}$ & 
              $+0.1688589 - 0.2257824 \, \mathrm{i}$ \\
    \midrule
    $A_{\lambda,\textrm{fin}}^{(1),\textrm{f},\textrm{ac}}
        /\textrm{GeV}^{-2}$ \hspace{-25pt} & 
        $\epsilon^0$ & $\epsilon^1$ & $\epsilon^2$ \\ 
    \midrule
    $(-,-)$ & $+0.0704773 + 0.1115518 \, \mathrm{i}$ & 
              $+0.1088695 + 0.1723194 \, \mathrm{i}$ &
              $+0.1505906 + 0.2383558 \, \mathrm{i}$ \\
    $(-,+)$ & $+0.0786529 - 0.1059797 \, \mathrm{i}$ & 
              $+0.1214989 - 0.1637118 \, \mathrm{i}$ & 
              $+0.1680598 - 0.2264496 \, \mathrm{i}$ \\
    $(+,-)$ & $+0.0700799 + 0.1118337 \, \mathrm{i}$ & 
              $+0.1082557 + 0.1727548 \, \mathrm{i}$ &
              $+0.1497415 + 0.2389581 \, \mathrm{i}$ \\
    $(+,+)$ & $+0.0790269 - 0.1056674 \, \mathrm{i}$ & 
              $+0.1220766 - 0.1632294 \, \mathrm{i}$ & 
              $+0.1688589 - 0.2257824 \, \mathrm{i}$ \\
    \bottomrule
\end{tabular}
\end{center}
\caption{Factorisable one-loop amplitudes at the point eq.~\eqref{eq:benchmark:point}, 
for the process eq.~\eqref{eq:process:physical}.}
\label{tab:benchmark:fact}
\end{table}

\begin{table}[!ht]
\begin{center}
\small
\begin{tabular}{cccc}
    \toprule
    $A_{\lambda, \textrm{fin}}^{(1), \textrm{nf}, \textrm{pr}}
        /\textrm{GeV}^{-2}$ \hspace{-25pt} & 
        $\epsilon^0$ & $\epsilon^1$ & $\epsilon^2$ \\
    \midrule
    $(-,-)$ & $+0.0749181 - 0.0466078 \, \mathrm{i}$ & 
              $+0.0218982 - 0.0146519 \, \mathrm{i}$ & 
              $-0.0683775 + 0.0392172 \, \mathrm{i}$ \\
    $(-,+)$ & $-0.0721223 - 0.0543451 \, \mathrm{i}$ & 
              $-0.0234780 - 0.0172730 \, \mathrm{i}$ & 
              $+0.0583115 + 0.0466539 \, \mathrm{i}$ \\
    $(+,-)$ & $+0.0768763 - 0.0474563 \, \mathrm{i}$ & 
              $+0.0246348 - 0.0157436 \, \mathrm{i}$ & 
              $-0.0646954 + 0.0369027 \, \mathrm{i}$ \\
    $(+,+)$ & $-0.0701521 - 0.0535693 \, \mathrm{i}$ & 
              $-0.0210973 - 0.0159031 \, \mathrm{i}$ & 
              $+0.0616122 + 0.0487801 \, \mathrm{i}$ \\
    \midrule
    $A_{\lambda,\textrm{fin}}^{(1),\textrm{nf},\textrm{ac}}
        /\textrm{GeV}^{-2}$ \hspace{-25pt} & 
        $\epsilon^0$ & $\epsilon^1$ & $\epsilon^2$ \\
    \midrule
    $(-,-)$ & $+0.0749181 - 0.0466078 \, \mathrm{i}$ & 
              $+0.0218982 - 0.0146519 \, \mathrm{i}$ & 
              $-0.0683775 + 0.0392172 \, \mathrm{i}$ \\
    $(-,+)$ & $-0.0721223 - 0.0543451 \, \mathrm{i}$ & 
              $-0.0234780 - 0.0172730 \, \mathrm{i}$ & 
              $+0.0583115 + 0.0466539 \, \mathrm{i}$ \\
    $(+,-)$ & $+0.0768763 - 0.0474563 \, \mathrm{i}$ & 
              $+0.0246348 - 0.0157436 \, \mathrm{i}$ & 
              $-0.0646954 + 0.0369027 \, \mathrm{i}$ \\
    $(+,+)$ & $-0.0701521 - 0.0535693 \, \mathrm{i}$ & 
              $-0.0210973 - 0.0159031 \, \mathrm{i}$ & 
              $+0.0616122 + 0.0487801 \, \mathrm{i}$ \\
    \bottomrule
\end{tabular}
\end{center}
\caption{Non-factorisable one-loop amplitudes at the point 
eq.~\eqref{eq:benchmark:point}, for the process 
eq.~\eqref{eq:process:physical}.}
\label{tab:benchmark:nfact}
\end{table}

\FloatBarrier


\bibliographystyle{JHEP}
\bibliography{literature.bib}

\providecommand{\href}[2]{#2}\begingroup\raggedright\begin{thebibliography}{10}

\bibitem{Ciccolini:2007ec}
M.~Ciccolini, A.~Denner and S.~Dittmaier, \emph{{Electroweak and QCD corrections to Higgs production via vector-boson fusion at the LHC}}, \href{https://doi.org/10.1103/PhysRevD.77.013002}{\emph{Phys. Rev. D} {\bfseries 77} (2008) 013002} [\href{https://arxiv.org/abs/0710.4749}{{\ttfamily 0710.4749}}].

\bibitem{Figy:2010ct}
T.~Figy, S.~Palmer and G.~Weiglein, \emph{{Higgs Production via Weak Boson Fusion in the Standard Model and the MSSM}}, \href{https://doi.org/10.1007/JHEP02(2012)105}{\emph{JHEP} {\bfseries 02} (2012) 105} [\href{https://arxiv.org/abs/1012.4789}{{\ttfamily 1012.4789}}].

\bibitem{Figy:2003nv}
T.~Figy, C.~Oleari and D.~Zeppenfeld, \emph{{Next-to-leading order jet distributions for Higgs boson production via weak boson fusion}}, \href{https://doi.org/10.1103/PhysRevD.68.073005}{\emph{Phys. Rev. D} {\bfseries 68} (2003) 073005} [\href{https://arxiv.org/abs/hep-ph/0306109}{{\ttfamily hep-ph/0306109}}].

\bibitem{Figy:2004pt}
T.~Figy and D.~Zeppenfeld, \emph{{QCD corrections to jet correlations in weak boson fusion}}, \href{https://doi.org/10.1016/j.physletb.2004.04.033}{\emph{Phys. Lett. B} {\bfseries 591} (2004) 297} [\href{https://arxiv.org/abs/hep-ph/0403297}{{\ttfamily hep-ph/0403297}}].

\bibitem{Berger:2004pca}
E.L.~Berger and J.M.~Campbell, \emph{{Higgs boson production in weak boson fusion at next-to-leading order}}, \href{https://doi.org/10.1103/PhysRevD.70.073011}{\emph{Phys. Rev. D} {\bfseries 70} (2004) 073011} [\href{https://arxiv.org/abs/hep-ph/0403194}{{\ttfamily hep-ph/0403194}}].

\bibitem{Bolzoni:2010xr}
P.~Bolzoni, F.~Maltoni, S.-O.~Moch and M.~Zaro, \emph{{Higgs production via vector-boson fusion at NNLO in QCD}}, \href{https://doi.org/10.1103/PhysRevLett.105.011801}{\emph{Phys. Rev. Lett.} {\bfseries 105} (2010) 011801} [\href{https://arxiv.org/abs/1003.4451}{{\ttfamily 1003.4451}}].

\bibitem{Bolzoni:2011cu}
P.~Bolzoni, F.~Maltoni, S.-O.~Moch and M.~Zaro, \emph{{Vector boson fusion at NNLO in QCD: SM Higgs and beyond}}, \href{https://doi.org/10.1103/PhysRevD.85.035002}{\emph{Phys. Rev. D} {\bfseries 85} (2012) 035002} [\href{https://arxiv.org/abs/1109.3717}{{\ttfamily 1109.3717}}].

\bibitem{Cacciari:2015jma}
M.~Cacciari, F.A.~Dreyer, A.~Karlberg, G.P.~Salam and G.~Zanderighi, \emph{{Fully Differential Vector-Boson-Fusion Higgs Production at Next-to-Next-to-Leading Order}}, \href{https://doi.org/10.1103/PhysRevLett.115.082002}{\emph{Phys. Rev. Lett.} {\bfseries 115} (2015) 082002} [\href{https://arxiv.org/abs/1506.02660}{{\ttfamily 1506.02660}}].

\bibitem{Cruz-Martinez:2018rod}
J.~Cruz-Martinez, T.~Gehrmann, E.W.N.~Glover and A.~Huss, \emph{{Second-order QCD effects in Higgs boson production through vector boson fusion}}, \href{https://doi.org/10.1016/j.physletb.2018.04.046}{\emph{Phys. Lett. B} {\bfseries 781} (2018) 672} [\href{https://arxiv.org/abs/1802.02445}{{\ttfamily 1802.02445}}].

\bibitem{Dreyer:2016oyx}
F.A.~Dreyer and A.~Karlberg, \emph{{Vector-Boson Fusion Higgs Production at Three Loops in QCD}}, \href{https://doi.org/10.1103/PhysRevLett.117.072001}{\emph{Phys. Rev. Lett.} {\bfseries 117} (2016) 072001} [\href{https://arxiv.org/abs/1606.00840}{{\ttfamily 1606.00840}}].

\bibitem{Liu:2019tuy}
T.~Liu, K.~Melnikov and A.A.~Penin, \emph{{Nonfactorizable QCD Effects in Higgs Boson Production via Vector Boson Fusion}}, \href{https://doi.org/10.1103/PhysRevLett.123.122002}{\emph{Phys. Rev. Lett.} {\bfseries 123} (2019) 122002} [\href{https://arxiv.org/abs/1906.10899}{{\ttfamily 1906.10899}}].

\bibitem{Long:2023mvc}
M.-M.~Long, K.~Melnikov and J.~Quarroz, \emph{{Non-factorizable virtual corrections to Higgs boson production in weak boson fusion beyond the eikonal approximation}}, \href{https://doi.org/10.1007/JHEP07(2023)035}{\emph{JHEP} {\bfseries 07} (2023) 035} [\href{https://arxiv.org/abs/2305.12937}{{\ttfamily 2305.12937}}].

\bibitem{Asteriadis:2023nyl}
K.~Asteriadis, C.~Br{\o}nnum-Hansen and K.~Melnikov, \emph{{Nonfactorizable corrections to Higgs boson production in weak boson fusion}}, \href{https://doi.org/10.1103/PhysRevD.109.014031}{\emph{Phys. Rev. D} {\bfseries 109} (2024) 014031} [\href{https://arxiv.org/abs/2305.08016}{{\ttfamily 2305.08016}}].

\bibitem{Berger:2019wnu}
N.~Berger et~al., \emph{{Simplified template cross sections {\textendash} Stage 1.1 and 1.2}}, \href{https://doi.org/10.21468/SciPostPhysCommRep.15}{\emph{SciPost Phys. Comm. Rep.} {\bfseries 15} (2026) } [\href{https://arxiv.org/abs/1906.02754}{{\ttfamily 1906.02754}}].

\bibitem{Dreyer:2020urf}
F.A.~Dreyer, A.~Karlberg and L.~Tancredi, \emph{{On the impact of non-factorisable corrections in VBF single and double Higgs production}}, \href{https://doi.org/10.1007/JHEP10(2020)131}{\emph{JHEP} {\bfseries 10} (2020) 131} [\href{https://arxiv.org/abs/2005.11334}{{\ttfamily 2005.11334}}].

\bibitem{Goetzfried:2026pbhb}
L.~G{\"o}tzfried and A.~von Manteuffel, \emph{{Two Feynman integral families for vector boson fusion at NNLO QCD}},  2026.

\bibitem{Chanowitz:1979zu}
M.S.~Chanowitz, M.~Furman and I.~Hinchliffe, \emph{{The Axial Current in Dimensional Regularization}}, \href{https://doi.org/10.1016/0550-3213(79)90333-X}{\emph{Nucl. Phys. B} {\bfseries 159} (1979) 225}.

\bibitem{Gehrmann:2015ora}
T.~Gehrmann, A.~von Manteuffel and L.~Tancredi, \emph{{The two-loop helicity amplitudes for $ q\overline{q}^{\prime}\to {V}_1{V}_2\to 4 $ leptons}}, \href{https://doi.org/10.1007/JHEP09(2015)128}{\emph{JHEP} {\bfseries 09} (2015) 128} [\href{https://arxiv.org/abs/1503.04812}{{\ttfamily 1503.04812}}].

\bibitem{Peraro:2020sfm}
T.~Peraro and L.~Tancredi, \emph{{Tensor decomposition for bosonic and fermionic scattering amplitudes}}, \href{https://doi.org/10.1103/PhysRevD.103.054042}{\emph{Phys. Rev. D} {\bfseries 103} (2021) 054042} [\href{https://arxiv.org/abs/2012.00820}{{\ttfamily 2012.00820}}].

\bibitem{Nogueira:1991ex}
P.~Nogueira, \emph{{Automatic Feynman Graph Generation}}, \href{https://doi.org/10.1006/jcph.1993.1074}{\emph{J. Comput. Phys.} {\bfseries 105} (1993) 279}.

\bibitem{Kuipers:2012rf}
J.~Kuipers, T.~Ueda, J.A.M.~Vermaseren and J.~Vollinga, \emph{{FORM version 4.0}}, \href{https://doi.org/10.1016/j.cpc.2012.12.028}{\emph{Comput. Phys. Commun.} {\bfseries 184} (2013) 1453} [\href{https://arxiv.org/abs/1203.6543}{{\ttfamily 1203.6543}}].

\bibitem{Davies:2026cci}
J.~Davies, T.~Kaneko, C.~Marinissen, T.~Ueda and J.A.M.~Vermaseren, \emph{{FORM Version 5.0}},  1, 2026.

\bibitem{tHooft:1972tcz}
G.~'t~Hooft and M.J.G.~Veltman, \emph{{Regularization and Renormalization of Gauge Fields}}, \href{https://doi.org/10.1016/0550-3213(72)90279-9}{\emph{Nucl. Phys. B} {\bfseries 44} (1972) 189}.

\bibitem{vanNeerven:1983vr}
W.L.~van Neerven and J.A.M.~Vermaseren, \emph{{Large loop integrals}}, \href{https://doi.org/10.1016/0370-2693(84)90237-5}{\emph{Phys. Lett. B} {\bfseries 137} (1984) 241}.

\bibitem{Mastrolia:2016dhn}
P.~Mastrolia, T.~Peraro and A.~Primo, \emph{{Adaptive Integrand Decomposition in parallel and orthogonal space}}, \href{https://doi.org/10.1007/JHEP08(2016)164}{\emph{JHEP} {\bfseries 08} (2016) 164} [\href{https://arxiv.org/abs/1605.03157}{{\ttfamily 1605.03157}}].

\bibitem{Goode:2024mci}
J.~Goode, F.~Herzog, A.~Kennedy, S.~Teale and J.~Vermaseren, \emph{{Tensor reduction for Feynman integrals with Lorentz and spinor indices}}, \href{https://doi.org/10.1007/JHEP11(2024)123}{\emph{JHEP} {\bfseries 11} (2024) 123} [\href{https://arxiv.org/abs/2408.05137}{{\ttfamily 2408.05137}}].

\bibitem{vonManteuffel:2025swv}
A.~von Manteuffel, D.~St{\"o}ckinger and M.~Wei{\ss}wange, \emph{{Four-loop renormalisation of chiral gauge theories with non-anticommuting {\ensuremath{\gamma}}$_{5}$ in the BMHV scheme}}, \href{https://doi.org/10.1007/JHEP08(2025)088}{\emph{JHEP} {\bfseries 08} (2025) 088} [\href{https://arxiv.org/abs/2506.12253}{{\ttfamily 2506.12253}}].

\bibitem{Cullen:2010jv}
G.~Cullen, M.~Koch-Janusz and T.~Reiter, \emph{{Spinney: A Form Library for Helicity Spinors}}, \href{https://doi.org/10.1016/j.cpc.2011.06.007}{\emph{Comput. Phys. Commun.} {\bfseries 182} (2011) 2368} [\href{https://arxiv.org/abs/1008.0803}{{\ttfamily 1008.0803}}].

\bibitem{Larin:1993tq}
S.A.~Larin, \emph{{The Renormalization of the axial anomaly in dimensional regularization}}, \href{https://doi.org/10.1016/0370-2693(93)90053-K}{\emph{Phys. Lett. B} {\bfseries 303} (1993) 113} [\href{https://arxiv.org/abs/hep-ph/9302240}{{\ttfamily hep-ph/9302240}}].

\bibitem{OlgosoRuiz:2024dzq}
P.~Olgoso~Ruiz and L.~Vecchi, \emph{{Spurious gauge-invariance and {\ensuremath{\gamma}}$_{5}$ in dimensional regularization}}, \href{https://doi.org/10.1007/JHEP12(2024)080}{\emph{JHEP} {\bfseries 12} (2024) 080} [\href{https://arxiv.org/abs/2406.17013}{{\ttfamily 2406.17013}}].

\bibitem{Ebert:2024xpy}
P.L.~Ebert, P.~K{\"u}hler, D.~St{\"o}ckinger and M.~Wei{\ss}wange, \emph{{Shedding light on evanescent shadows {\textemdash} Exploration of non-anticommuting {\ensuremath{\gamma}}$_{5}$ in Dimensional Regularisation}}, \href{https://doi.org/10.1007/JHEP01(2025)114}{\emph{JHEP} {\bfseries 01} (2025) 114} [\href{https://arxiv.org/abs/2411.02543}{{\ttfamily 2411.02543}}].

\bibitem{Kuhler:2025znv}
P.~K{\"u}hler and D.~St{\"o}ckinger, \emph{{Two-loop renormalization of a chiral SU(2) gauge theory in dimensional regularization with non-anticommuting {\ensuremath{\gamma}}$_{5}$}}, \href{https://doi.org/10.1007/JHEP07(2025)082}{\emph{JHEP} {\bfseries 07} (2025) 082} [\href{https://arxiv.org/abs/2504.06080}{{\ttfamily 2504.06080}}].

\bibitem{Chetyrkin:1981qh}
K.G.~Chetyrkin and F.V.~Tkachov, \emph{{Integration by parts: The algorithm to calculate $\beta$-functions in 4 loops}}, \href{https://doi.org/10.1016/0550-3213(81)90199-1}{\emph{Nucl. Phys. B} {\bfseries 192} (1981) 159}.

\bibitem{Laporta:2000dsw}
S.~Laporta, \emph{{High-precision calculation of multiloop Feynman integrals by difference equations}}, \href{https://doi.org/10.1142/S0217751X00002159}{\emph{Int. J. Mod. Phys. A} {\bfseries 15} (2000) 5087} [\href{https://arxiv.org/abs/hep-ph/0102033}{{\ttfamily hep-ph/0102033}}].

\bibitem{vonManteuffel:2012np}
A.~von Manteuffel and C.~Studerus, \emph{{Reduze 2 - Distributed Feynman Integral Reduction}},  1, 2012.

\bibitem{tHooft:1978jhc}
G.~'t~Hooft and M.J.G.~Veltman, \emph{{Scalar One Loop Integrals}}, \href{https://doi.org/10.1016/0550-3213(79)90605-9}{\emph{Nucl. Phys. B} {\bfseries 153} (1979) 365}.

\bibitem{Denner:1999gp}
A.~Denner, S.~Dittmaier, M.~Roth and D.~Wackeroth, \emph{{Predictions for all processes e+ e- ---{\ensuremath{>}} 4 fermions + gamma}}, \href{https://doi.org/10.1016/S0550-3213(99)00437-X}{\emph{Nucl. Phys. B} {\bfseries 560} (1999) 33} [\href{https://arxiv.org/abs/hep-ph/9904472}{{\ttfamily hep-ph/9904472}}].

\bibitem{Kotikov:1990kg}
A.V.~Kotikov, \emph{{Differential equations method: New technique for massive Feynman diagrams calculation}}, \href{https://doi.org/10.1016/0370-2693(91)90413-K}{\emph{Phys. Lett. B} {\bfseries 254} (1991) 158}.

\bibitem{Henn:2013pwa}
J.M.~Henn, \emph{{Multiloop integrals in dimensional regularization made simple}}, \href{https://doi.org/10.1103/PhysRevLett.110.251601}{\emph{Phys. Rev. Lett.} {\bfseries 110} (2013) 251601} [\href{https://arxiv.org/abs/1304.1806}{{\ttfamily 1304.1806}}].

\bibitem{Henn:2020lye}
J.~Henn, B.~Mistlberger, V.A.~Smirnov and P.~Wasser, \emph{{Constructing d-log integrands and computing master integrals for three-loop four-particle scattering}}, \href{https://doi.org/10.1007/JHEP04(2020)167}{\emph{JHEP} {\bfseries 04} (2020) 167} [\href{https://arxiv.org/abs/2002.09492}{{\ttfamily 2002.09492}}].

\bibitem{Baikov:1996iu}
P.A.~Baikov, \emph{{Explicit solutions of the multiloop integral recurrence relations and its application}}, \href{https://doi.org/10.1016/S0168-9002(97)00126-5}{\emph{Nucl. Instrum. Meth. A} {\bfseries 389} (1997) 347} [\href{https://arxiv.org/abs/hep-ph/9611449}{{\ttfamily hep-ph/9611449}}].

\bibitem{Abreu:2018rcw}
S.~Abreu, B.~Page and M.~Zeng, \emph{{Differential equations from unitarity cuts: nonplanar hexa-box integrals}}, \href{https://doi.org/10.1007/JHEP01(2019)006}{\emph{JHEP} {\bfseries 01} (2019) 006} [\href{https://arxiv.org/abs/1807.11522}{{\ttfamily 1807.11522}}].

\bibitem{Jiang:2024eaj}
X.~Jiang, J.~Liu, X.~Xu and L.L.~Yang, \emph{{Symbol letters of Feynman integrals from Gram determinants}}, \href{https://doi.org/10.1016/j.physletb.2025.139443}{\emph{Phys. Lett. B} {\bfseries 864} (2025) 139443} [\href{https://arxiv.org/abs/2401.07632}{{\ttfamily 2401.07632}}].

\bibitem{Dlapa:2023cvx}
C.~Dlapa, M.~Helmer, G.~Papathanasiou and F.~Tellander, \emph{{Symbol alphabets from the Landau singular locus}}, \href{https://doi.org/10.1007/JHEP10(2023)161}{\emph{JHEP} {\bfseries 10} (2023) 161} [\href{https://arxiv.org/abs/2304.02629}{{\ttfamily 2304.02629}}].

\bibitem{Heller:2019gkq}
M.~Heller, A.~von Manteuffel and R.M.~Schabinger, \emph{{Multiple polylogarithms with algebraic arguments and the two-loop EW-QCD Drell-Yan master integrals}}, \href{https://doi.org/10.1103/PhysRevD.102.016025}{\emph{Phys. Rev. D} {\bfseries 102} (2020) 016025} [\href{https://arxiv.org/abs/1907.00491}{{\ttfamily 1907.00491}}].

\bibitem{Matijasic:2024too}
A.~Matija{\v{s}}i{\'c}, \emph{{Singularity structure of Feynman integrals with applications to six-particle scattering processes}}, Ph.D. thesis, Munich U., 2024.
\newblock 10.5282/edoc.34154.

\bibitem{Czakon:2026tog}
M.~Czakon and L.~Tancredi, \emph{{Solution of Canonical Differential Equations for Integrals on Arbitrary Geometries}},  \href{https://arxiv.org/abs/2606.30354}{{\ttfamily 2606.30354}}.

\bibitem{Gehrmann:2018yef}
T.~Gehrmann, J.M.~Henn and N.A.~Lo~Presti, \emph{{Pentagon functions for massless planar scattering amplitudes}}, \href{https://doi.org/10.1007/JHEP10(2018)103}{\emph{JHEP} {\bfseries 10} (2018) 103} [\href{https://arxiv.org/abs/1807.09812}{{\ttfamily 1807.09812}}].

\bibitem{Chicherin:2020oor}
D.~Chicherin and V.~Sotnikov, \emph{{Pentagon Functions for Scattering of Five Massless Particles}}, \href{https://doi.org/10.1007/JHEP12(2020)167}{\emph{JHEP} {\bfseries 20} (2020) 167} [\href{https://arxiv.org/abs/2009.07803}{{\ttfamily 2009.07803}}].

\bibitem{Chicherin:2021dyp}
D.~Chicherin, V.~Sotnikov and S.~Zoia, \emph{{Pentagon functions for one-mass planar scattering amplitudes}}, \href{https://doi.org/10.1007/JHEP01(2022)096}{\emph{JHEP} {\bfseries 01} (2022) 096} [\href{https://arxiv.org/abs/2110.10111}{{\ttfamily 2110.10111}}].

\bibitem{Abreu:2023rco}
S.~Abreu, D.~Chicherin, H.~Ita, B.~Page, V.~Sotnikov, W.~Tschernow et~al., \emph{{All Two-Loop Feynman Integrals for Five-Point One-Mass Scattering}}, \href{https://doi.org/10.1103/PhysRevLett.132.141601}{\emph{Phys. Rev. Lett.} {\bfseries 132} (2024) 141601} [\href{https://arxiv.org/abs/2306.15431}{{\ttfamily 2306.15431}}].

\bibitem{Badger:2025ljy}
S.~Badger, M.~Becchetti, C.~Brancaccio, M.~Czakon, H.B.~Hartanto, R.~Poncelet et~al., \emph{{Double virtual QCD corrections to $ t\overline{t} $+jet production at the LHC}}, \href{https://doi.org/10.1007/JHEP05(2026)044}{\emph{JHEP} {\bfseries 05} (2026) 044} [\href{https://arxiv.org/abs/2511.11424}{{\ttfamily 2511.11424}}].

\bibitem{Goncharov:2005sla}
A.B.~Goncharov, \emph{{Galois symmetries of fundamental groupoids and noncommutative geometry}}, \href{https://doi.org/10.1215/S0012-7094-04-12822-2}{\emph{Duke Math. J.} {\bfseries 128} (2005) 209} [\href{https://arxiv.org/abs/math/0208144}{{\ttfamily math/0208144}}].

\bibitem{Goncharov:2010jf}
A.B.~Goncharov, M.~Spradlin, C.~Vergu and A.~Volovich, \emph{{Classical Polylogarithms for Amplitudes and Wilson Loops}}, \href{https://doi.org/10.1103/PhysRevLett.105.151605}{\emph{Phys. Rev. Lett.} {\bfseries 105} (2010) 151605} [\href{https://arxiv.org/abs/1006.5703}{{\ttfamily 1006.5703}}].

\bibitem{Duhr:2011zq}
C.~Duhr, H.~Gangl and J.R.~Rhodes, \emph{{From polygons and symbols to polylogarithmic functions}}, \href{https://doi.org/10.1007/JHEP10(2012)075}{\emph{JHEP} {\bfseries 10} (2012) 075} [\href{https://arxiv.org/abs/1110.0458}{{\ttfamily 1110.0458}}].

\bibitem{Duhr:2012fh}
C.~Duhr, \emph{{Hopf algebras, coproducts and symbols: an application to Higgs boson amplitudes}}, \href{https://doi.org/10.1007/JHEP08(2012)043}{\emph{JHEP} {\bfseries 08} (2012) 043} [\href{https://arxiv.org/abs/1203.0454}{{\ttfamily 1203.0454}}].

\bibitem{Pari:2025}
{The {PARI}~{G}roup}, \emph{{PARI/GP, {V}ersion 2.17.2}}.
\newblock Univ. Bordeaux, 2025.

\bibitem{Boost:2026}
{The Boost Community}, \emph{Boost {C}++ {L}ibraries, {V}ersion 1.83}, 2025.

\bibitem{Hidding:2020ytt}
M.~Hidding, \emph{{DiffExp, a Mathematica package for computing Feynman integrals in terms of one-dimensional series expansions}}, \href{https://doi.org/10.1016/j.cpc.2021.108125}{\emph{Comput. Phys. Commun.} {\bfseries 269} (2021) 108125} [\href{https://arxiv.org/abs/2006.05510}{{\ttfamily 2006.05510}}].

\bibitem{flint:2025}
{The {FLINT} team}, \emph{{FLINT}: {F}ast {L}ibrary for {N}umber {T}heory, {V}ersion 3.4.0}, 2025.

\bibitem{Giele:1991vf}
W.T.~Giele and E.W.N.~Glover, \emph{{Higher Order Corrections to Jet Cross Sections in e+ e- Annihilation}}, \href{https://doi.org/10.1103/PhysRevD.46.1980}{\emph{Phys. Rev. D} {\bfseries 46} (1992) 1980}.

\bibitem{Kunszt:1994np}
Z.~Kunszt, A.~Signer and Z.~Trocsanyi, \emph{{Singular terms of helicity amplitudes at one loop in QCD and the soft limit of the cross-sections of multiparton processes}}, \href{https://doi.org/10.1016/0550-3213(94)90077-9}{\emph{Nucl. Phys. B} {\bfseries 420} (1994) 550} [\href{https://arxiv.org/abs/hep-ph/9401294}{{\ttfamily hep-ph/9401294}}].

\bibitem{Catani:1996vz}
S.~Catani and M.H.~Seymour, \emph{{A general algorithm for calculating jet cross-sections in NLO QCD}}, \href{https://doi.org/10.1016/S0550-3213(96)00589-5}{\emph{Nucl. Phys. B} {\bfseries 485} (1997) 291} [\href{https://arxiv.org/abs/hep-ph/9605323}{{\ttfamily hep-ph/9605323}}].

\bibitem{Catani:1996mi}
S.~Catani and M.H.~Seymour, \emph{{QCD jet calculations in DIS based on the subtraction method and dipole formalism}},  in \emph{{4th International Workshop on Deep Inelastic Scattering and Related Phenomena}}, pp.~454--458, 9, 1996 [\href{https://arxiv.org/abs/hep-ph/9609237}{{\ttfamily hep-ph/9609237}}].

\bibitem{Becher:2009cu}
T.~Becher and M.~Neubert, \emph{{Infrared singularities of scattering amplitudes in perturbative QCD}}, \href{https://doi.org/10.1103/PhysRevLett.102.162001}{\emph{Phys. Rev. Lett.} {\bfseries 102} (2009) 162001} [\href{https://arxiv.org/abs/0901.0722}{{\ttfamily 0901.0722}}].

\bibitem{Heller:2021qkz}
M.~Heller and A.~von Manteuffel, \emph{{MultivariateApart: Generalized partial fractions}}, \href{https://doi.org/10.1016/j.cpc.2021.108174}{\emph{Comput. Phys. Commun.} {\bfseries 271} (2022) 108174} [\href{https://arxiv.org/abs/2101.08283}{{\ttfamily 2101.08283}}].

\bibitem{Borowka:2017idc}
S.~Borowka, G.~Heinrich, S.~Jahn, S.P.~Jones, M.~Kerner, J.~Schlenk et~al., \emph{{pySecDec: A toolbox for the numerical evaluation of multi-scale integrals}}, \href{https://doi.org/10.1016/j.cpc.2017.09.015}{\emph{Comput. Phys. Commun.} {\bfseries 222} (2018) 313} [\href{https://arxiv.org/abs/1703.09692}{{\ttfamily 1703.09692}}].

\bibitem{Hahn:2000jm}
T.~Hahn, \emph{{Automatic loop calculations with FeynArts, FormCalc, and LoopTools}}, \href{https://doi.org/10.1016/S0920-5632(00)00848-3}{\emph{Nucl. Phys. B Proc. Suppl.} {\bfseries 89} (2000) 231} [\href{https://arxiv.org/abs/hep-ph/0005029}{{\ttfamily hep-ph/0005029}}].

\bibitem{Campanario:2013fsa}
F.~Campanario, T.M.~Figy, S.~Pl{\"a}tzer and M.~Sj{\"o}dahl, \emph{{Electroweak Higgs Boson Plus Three Jet Production at Next-to-Leading-Order QCD}}, \href{https://doi.org/10.1103/PhysRevLett.111.211802}{\emph{Phys. Rev. Lett.} {\bfseries 111} (2013) 211802} [\href{https://arxiv.org/abs/1308.2932}{{\ttfamily 1308.2932}}].

\end{thebibliography}\endgroup

\end{document}